\documentclass{article}
\usepackage{PRIMEarxiv}

\usepackage{hyperref}
\usepackage{url}            
\usepackage{booktabs}       
\usepackage{amsfonts}       
\usepackage{nicefrac}       
\usepackage{microtype}      
\usepackage{lipsum}
\usepackage{fancyhdr}       
\usepackage{graphicx}       
\graphicspath{{media/}}     
\usepackage{cite}
\usepackage{tikz}
\usetikzlibrary{shapes.geometric, arrows.meta, positioning, calc, shadows, patterns, decorations.pathreplacing}
\usepackage{amsmath,amssymb,amsfonts}
\usepackage{graphicx}
\usepackage{algorithmicx}
\usepackage{textcomp}
\usepackage{mathtools}
\usepackage{xcolor}
\def\BibTeX{{\rm B\kern-.05em{\sc i\kern-.025em b}\kern-.08em
    T\kern-.1667em\lower.7ex\hbox{E}\kern-.125emX}}

\usepackage{multirow}
\usepackage{soul,xcolor}
\usepackage{comment}
\usepackage{makecell}
\usepackage{amsmath}
\usepackage{amsmath, amssymb, amscd, amsthm, amsfonts}
\usepackage{colortbl}
\usepackage{soul}
\usepackage{changepage}
\usepackage{algorithm2e}

\usepackage{algpseudocode}
\usepackage{subcaption}
\usepackage{multirow}
\usepackage{mdframed}
\mdfdefinestyle{custombox}{
    linecolor=blue,        
    linewidth=2pt,         
    roundcorner=10pt,      
    backgroundcolor=blue!5, 
    innertopmargin=10pt,
    innerbottommargin=10pt,
    innerleftmargin=10pt,
    innerrightmargin=10pt
}
\usepackage{etoolbox} 

\definecolor{grayheader}{RGB}{0, 0, 139}
\definecolor{cadetblue}{RGB}{139, 0, 0}

\newmdenv[
  linewidth=1.2pt,
  linecolor=grayheader,
  roundcorner=6pt,
  backgroundcolor=white,
  innertopmargin=1em,
  innerbottommargin=1em,
  innerleftmargin=1em,
  innerrightmargin=1em,
  frametitlebackgroundcolor=grayheader,
  frametitlefont=\color{white}\bfseries,
  frametitle=Prompt
]{promptbox}

\newmdenv[
  linewidth=1.2pt,
  linecolor=cadetblue,
  roundcorner=6pt,
  backgroundcolor=white,
  innertopmargin=1em,
  innerbottommargin=1em,
  innerleftmargin=1em,
  innerrightmargin=1em,
  frametitlebackgroundcolor=cadetblue,
  frametitlefont=\color{white}\bfseries,
  frametitle=Response
]{responsebox}
\title{Generative AI for Critical Infrastructure in Smart Grids: A Unified Framework for Synthetic Data Generation and Anomaly Detection
}

\author{
  Aydin Zaboli and Junho Hong\\
  University of Michigan-Dearborn \\
  Dearborn, MI United States\\
  \texttt{\{azaboli, jhwr\}@umich.edu} \\
}

\begin{document}
\maketitle

\begin{abstract}
In digital substations, security events pose significant challenges to the sustained operation of power systems. To mitigate these challenges, the implementation of robust defense strategies is critically important. A thorough process of anomaly identification and detection in information and communication technology (ICT) frameworks is crucial to ensure secure and reliable communication and coordination between interconnected devices within digital substations. Hence, this paper addresses the critical cybersecurity challenges confronting IEC61850-based digital substations within modern smart grids, where the integration of advanced communication protocols, e.g., generic object-oriented substation event (GOOSE), has enhanced energy management and introduced significant vulnerabilities to cyberattacks. Focusing on the limitations of traditional anomaly detection systems (ADSs) in detecting threats, this research proposes a transformative approach by leveraging generative AI (GenAI) to develop robust ADSs. The primary contributions include the suggested advanced adversarial traffic mutation (AATM) technique to generate synthesized and balanced datasets for GOOSE messages, ensuring protocol compliance and enabling realistic zero-day attack pattern creation to address data scarcity. Then, the implementation of GenAI-based ADSs incorporating the task-oriented dialogue (ToD) processes has been explored for improved detection of attack patterns. Finally, a comparison of the GenAI-based ADS with machine learning (ML)-based ADSs has been implemented to showcase the outperformance of the GenAI-based frameworks considering the AATM-generated GOOSE datasets and standard/advanced performance evaluation metrics.
\end{abstract}

\newtheorem{theorem}{Theorem}
\newtheorem{lemma}[theorem]{Lemma}
\newtheorem{proposition}[theorem]{Proposition}
\newtheorem{definition}[theorem]{Definition}
\mdfdefinestyle{abstractbox}{
  backgroundcolor=gray!10,
  linecolor=white,
  linewidth=0.5pt,
  roundcorner=5pt,
  innertopmargin=10pt,
  innerbottommargin=10pt,
  innerleftmargin=10pt,
  innerrightmargin=10pt,
  skipabove=12pt,
  skipbelow=12pt
}

\section{Introduction} 
Power grid blackouts considerably interrupt societal and economic operations, caused by factors such as human errors, technical malfunctions, and environmental incidents. Additionally, the risk of cyberattacks inducing these outages highlights an increasing susceptibility. Consequently, ensuring the cybersecurity of ICT systems integral to power grid functions has emerged as an imperative need~\cite{hong2014detection}. Digital substations utilizing IEC61850 are pivotal within the power grid architecture, overseeing the allocation, conversion, and integration of energy streams. The advent of smart grids has amalgamated the power grid infrastructure with communication networks and computing functions, paving the way for numerous groundbreaking applications such as automated data acquisition and the remote management of electrical systems and elements~\cite{sun2018cyber, hong2022automated}. Nevertheless, integrating these systems introduces a range of security vulnerabilities to the smart grid. ADSs play a crucial role in identifying and mitigating malicious activities by adversaries. Historically, ADSs have proven effective in traditional ICT fields for this purpose. However, with the adoption of IEC61850 and the implementation of specific communication protocols, such as multicast messages (e.g., GOOSE), new avenues have emerged for customized malicious strategies that exhibit distinct traffic and attack patterns. These may comprise unauthorized data interceptions and denial-of-service (DoS) attacks. Thus, it is essential for ADSs to develop and refine new signatures for training, testing, validation, and assessment to address these emerging challenges effectively~\cite{10339874, zaboli2025advanced}. 
\subsection{Problem Statement}
The criticality of substations is grounded in their function as nodes that handle multiple transmission lines, a feature that amplifies their impact on grid stability. Traditional contingency planning, such as $N-1$ analyses, assesses the system's resilience to the loss of a single component, but the compromise of a substation due to different types of cyberattack can lead to an $N-m$ scenario—simultaneous outages of multiple lines—overwhelming the grid's capacity and triggering catastrophic disruptions. This vulnerability is particularly heightened in the U.S., where the expansion of unmanned substations, dependent on remote access for maintenance, has created significant exposure points. These facilities are ideal targets for cyber intrusions, as both authorized engineers and malicious actors can exploit the same access mechanisms. The reliance on remote access in unmanned substations has advantages and limitations; while it facilitates operational efficiency, it also increases cybersecurity risks. The potential for unauthorized access by intruders, leveraging the same entry points as legitimate operators, underscores a critical gap in current security frameworks. This issue is compounded by the evolving nature of cyber threats, which can exploit vulnerabilities in communication protocols, leading to scenarios such as false data injection (FDI), DoS, and replay (RE) attacks. The need to reinforce these systems is evident, given their role in ensuring grid reliability and the potential for cascading failures that can disrupt entire regions.

Furthermore, ML techniques applied within ADSs play a crucial role in detecting and correcting inconsistencies in GOOSE multicast transmissions. These techniques are acknowledged for their accuracy and emphasis on data, offering a sophisticated foundation for cybersecurity measures. Nonetheless, they do present certain obstacles. A significant limitation is the need for ongoing model retraining whenever new attack vectors emerge. Upon detection of a novel attack pattern, the ML models require updating to integrate this novel data. This retraining procedure demands substantial time and resources, which results in a period of vulnerability during which the system is exposed to new threats (e.g., zero-day attacks) that have not yet been included in the model's intelligence framework~\cite{beg2023review}. Furthermore, the ability of these ML-driven ADSs to scale, alongside their efficiency in decision-making and data processing, is critically significant for the operational functionality. Scalability concerns focus on the model's proficiency in adapting and sustaining performance as the network expands or as data volume increases. The decision-making aspect relates to the model's capability for accurately identifying secure versus malicious actions, a challenge that grows more complicated with the advancement of sophisticated attack methods. Finally, the domain of data processing highlights the necessity for proficient management and analysis of extensive datasets that are frequently essential for system operations~\cite{hong2017intelligent, chen2016modeling}. The identified key deficiencies in existing ML-based ADSs pose substantial difficulties within industrial control system (ICS) networks, especially in settings where GOOSE messages are fundamental to time-sensitive communications.
\subsection{Research Objectives}
The successful fusion of traditional electrical grids with advanced communication networks and computational frameworks through smart grid technologies brings about significant security risks, particularly through the IEC61850 standard and its related protocols. This integration facilitates unique attack vectors marked by distinct traffic and attack patterns that pose considerable challenges to conventional ADSs. Addressing this evolving security framework requires the development of enhanced ADS capabilities featuring specialized signatures for robust training, testing, and validation against novel threats. This is especially crucial for identifying zero-day attacks in traffic patterns, which lack predetermined rules in practical scenarios, further complicated by the scarcity of realistic, balanced, and comprehensive IEC61850-based communication datasets. In light of these significant challenges, it is essential to develop adaptive, resilient, and scalable AD solutions that effectively reduce latency in integrating new threat intelligence while simultaneously improving decision-making processes and data management capabilities. Applications or systems known as GenAI tools employ large language models (LLMs) and sometimes other AI models to produce content customized to user inputs within distinct contexts, presenting a revolutionary approach via platforms such as OpenAI's ChatGPT~\cite{OpenAIChatGPT}, Anthropic Claude Pro~\cite{anthropic} and Microsoft Copilot AI~\cite{Copilot}. These GenAI solutions, crafted with precision for deep contextual understanding using advanced memory architectures and natural language processing (NLP) capabilities, demonstrate the ability to identify zero-day attacks through contextual evaluation, even in the absence of substantial prior knowledge of particular threat signatures~\cite{smolin2024gencoder}, thus significantly decreasing the workload on human operators relative to traditional approaches that rely extensively on routine retraining procedures and fixed data models. Synthesizing pre-processed datasets and engineering ADSs that integrate past trends while independently analyzing changing network conditions and suggesting responses based on data insights, the goal of this research is to design security frameworks that possess the ability to autonomously respond, continuously learn, and process data at scale, thereby improving the robustness and dependability of digital substations. This is achieved by ensuring that innovative or complex threats are promptly detected, preventing significant disruptions to essential infrastructure systems~\cite{zaboli2024chatgpt, hong2022automated, zaboli2024leveraging, gill2023chatgpt, ten2011anomaly}.
\subsection{Literature Review} \label{lit_survey}
The evolution of digital substations has necessitated the development of sophisticated ADSs leveraging ML techniques. These systems analyze data patterns to identify cyber threats in real-time, ensuring power grid reliability~\cite{choi2020multi, kreimel2020anomaly, wang2022anomaly}. Recent research has explored diverse ML approaches for AD. Alvee~\textit{et al.}~\cite{alvee2021ransomware} developed a convolutional neural network (CNN)-based methodology that converts binary files into images for ransomware detection, though limited by dataset constraints and scenario coverage. A real-time ADS using advanced ML was explored in~\cite{panthi2020anomaly}, but faced challenges in processing large data volumes efficiently. Hybrid approaches combining CNN and long short-term memory (LSTM) networks~\cite{ankitdeshpandey2020development} showed promise but struggled with generalization to new attack types. Several studies focused on IEC61850 protocol-specific solutions. Eynawi~\textit{et al.}~\cite{eynawi2024machine} developed ML-based feature selection for GOOSE and Sampled Value (SV) messages, though computational overhead limited real-time deployment. Quincozes~\textit{et al.}~\cite{quincozes2022feature} introduced innovative feature engineering for IEC61850, but relied on static datasets. A game theory integration by Jay~\cite{jay2023deception} offered proactive defense mechanisms, though practical implementation remained computationally intensive. Bhattacharya~\textit{et al.}~\cite{bhattacharya2024ml} achieved impressive results with k-Nearest Neighbors (KNN), but lacked deep learning (DL) integration and real-world deployment validation. Other approaches included gradient boosting~\cite{upadhyay2020gradient}, distributed security systems~\cite{zhu2020intrusion}, and ML-driven GOOSE message analysis~\cite{ustun2021machine}, each with specific limitations in scalability, computational demands, or protocol coverage. A GenAI-based ADS considering the human-in-the-loop (HITL) was proposed by Zaboli~\textit{et al.}~\cite{zaboli2024chatgpt} to implement the detection processes in IEC61850-based multicast messages, considering different GPT tools. However, there was no continuous learning and automated processing of actions in this research alongside the multicast messages extracted from the hardware-in-the-loop (HIL) testbed. Moreover, Zaboli~\textit{et al.}~\cite{zaboli2025advanced} suggested a novel GenAI-based ToD framework for the AD process to overcome the challenges given in the GenAI-based ADS using the HITL. It covered the learning and automated processes gaps and made a good comparison with the HITL technique considering different GPT tools. While this framework showcased the outperformance over the HITL process, the lack of a comparison of this framework with ML-based ADS was a gap of the research. Also, the balance and realistic issues of datasets considering the zero-day attack were missed as new threats emerged in real-world applications.

Recent data balancing methodologies build upon and extend established traditional approaches. Bhattacharya \textit{et al.}~\cite{bhattacharya2024ml} employed the RUS and SMOTE for imbalanced datasets in IEC61850-based messages. Although SMOTE can efficiently generate new samples for datasets with few dimensions, its performance drops markedly when dealing with high-dimensional data. Moreover, SMOTE's method of interpolation does not always guarantee that the synthetic data will be of high quality or that the interpolation is performed in the best possible way, which can lead to considerable noise in the dataset. Recent developments in GANs have achieved outstanding results in computer vision and image synthesis. As a result, researchers are now exploring the use of GANs to generate samples for minority classes, with the aim of addressing data imbalance challenges~\cite{yuan2023data}.
Some challenges in the data pre-processing part for the AD process can include data imbalance, absence of prior information, increased data complexity, data generation for anomalies with different patterns, resource extensiveness, and re-training process for zero-day attacks, which can be found in Table~\ref{tab:prep-challenges} with the relevant descriptions for each challenge~\cite{dromard2020study, lin2019dynamic, mbona2022detecting, yaacoub2020cyber, fu2022deep, bhattacharya2024ml, boukerche2020outlier}.
\begin{table*}[htbp]
\centering
\caption{The challenges encountered in detecting anomalies within network systems in terms of data availability.}
\label{tab:prep-challenges}
\begin{tabular}{|p{5.5cm}|p{10cm}|}
\hline
\makecell{\textbf{Challenge}} & \makecell{\textbf{Description}} \\
\hline
Excessive dependence on prior knowledge~\cite{dromard2020study}
 & Contemporary approaches to AD exhibit a pronounced reliance on established attack signatures and predefined network behavioral baselines. Such dependence on historical data and fixed analytical frameworks can restrict their effectiveness in detecting novel or rapidly evolving security threats (zero-day attacks). \\
 \hline
Resource-demanding \& time-consuming implementation~\cite{lin2019dynamic}
 & Generating new signatures or updating profiles for current detection systems entails considerable time and resource investment, typically requiring the specialized expertise of network security professionals. \\
 \hline
Unavailability of prior contextual attack data~\cite{mbona2022detecting}
 & The scarcity of preceding intelligence, particularly regarding threat vectors such as zero-day attacks, represents a significant operational challenge. Moreover, unknown system vulnerabilities and evolving attack methodologies often further complicate the implementation of effective cybersecurity countermeasures. \\
 \hline
Elevated data complexity and absence of real-time detection mechanisms~\cite{yaacoub2020cyber}
 & Due to the exponential increase in data complexity and size, network traffic continues to intensify, making the real-time attack detection and the maintenance of consistent monitoring increasingly difficult. \\
 \hline
Data augmentation aimed at anomalous trend emergence~\cite{fu2022deep}
 & Conventional AD methods often encounter difficulties in producing unknown anomalous datasets, which are crucial for effectively training and evaluating detection algorithms. \\
\hline
Imbalanced class~\cite{bhattacharya2024ml, boukerche2020outlier}
 & The network AD inherently suffers from class imbalance, characterized by a privilege of normal instances relative to a scarcity of abnormal ones. Consequently, AD models may experience performance degradation when faced with extreme discrepancies in training samples, underscoring the criticality of addressing data imbalance. \\
 \hline
\end{tabular}
\end{table*}
Dairi \textit{et al.} introduced semi‐supervised DL schemes that effectively learn temporal dependencies using only normal training data; nonetheless, their dependence on IEC60870-5-104 datasets (for SCADA control) creates a challenge when extending these methods to the semantics of GOOSE messages in digital substations~\cite{dairi2023semi}. López \textit{et al.}~\cite{lopez2022substation} developed a substation-aware ADS that integrates substation topology to enhance detection accuracy, although its limited capability to infer the deep semantic content of GOOSE messages remains a critical gap. Moreover, Sahani \textit{et al.}~\cite{sahani2023machine} provided a comprehensive survey of ML-based ADS in smart grids, offering valuable insights into ML applications while also revealing scalability and real-time processing challenges. Additionally, Anwar \textit{et al.}~\cite{anwar2021comparison} compared unsupervised learning algorithms for intrusion detection in the IEC60870-5-104 SCADA protocol, contributing to the understanding of detection performance while exposing low detection accuracies that limit practical deployment. An ADS along with network packet features for wide-area protection was proposed by Singh and Govindarasu~\cite{singh2021cyber}, though the complexity in feature selection and high processing overhead hinders a smooth integration. Khaw \textit{et al.}~\cite{khaw2020deep} presented a universal DL-based cyberattack detection system for transmission protective relays that simplifies model tuning across different fault types, but its reliance on static training datasets limits its adaptability to evolving zero-day attacks.
Lim \textit{et al.}~\cite{lim2024future} reviewed the application of GANs for AD in network security, demonstrating that these models can generate synthetic minority samples to improve the detection of rare attack patterns; however, they also highlighted challenges such as an over-reliance on predefined evaluation metrics and insufficient representation learning when dealing with extremely imbalanced datasets. Similarly, Yuan \textit{et al.}~\cite{yuan2023data} proposed a GAN framework, which leverages LSTM networks to capture temporal features and generate high-quality synthetic intrusion data for industrial control systems, enhancing the AD performance on imbalanced datasets; however, their approach is challenged by issues of mode collapse and limited diversity among the generated anomalies. In addition, Sauber‑Cole \textit{et al.}~\cite{sauber2022use} surveyed GAN techniques for mitigating class imbalance in tabular data, emphasizing the promise of GAN architectures to produce representative minority instances; nevertheless, they noted persistent challenges, including architectural sensitivity and the lack of standardized evaluation metrics to reliably assess synthetic data quality. 
Manzoor \textit{et al.} introduced a method that exploits the in‑context learning (ICL) capability inherent in transformer architectures to detect zero‑day attacks in digital substations. Their technique allows the model to integrate new attack examples with minimal retraining, achieving detection accuracies exceeding 85\% for zero‑day scenarios where conventional state‑of‑the‑art baselines fail. Nonetheless, the method is contingent on the diversity of training data, as its efficacy is highly dependent on the number and heterogeneity of attack classes included during training. Moreover, the overall performance of the system is significantly influenced by the quality of the weak classifiers, with notable performance degradation observed. Further, this research used the multi-mixing technique to generate synthetic datasets without good validation and meeting the IEC61850-based messages violation rules, which is a critical gap~\cite{manzoor2025detecting, manzoor2024zero}. Lin \textit{et al.}~\cite{2024_3C_causalprompt} developed ``CausalPrompt,'' a novel prompting strategy designed to adapt LLMs for classification and regression tasks via weakly supervised causal reasoning. By integrating domain-specific causal inferences during the fine‑tuning phase, their approach enhanced the adaptability and resilience of energy systems against data distribution shifts. Their experimental results revealed improvements in predictive performance under feature changes. However, the method encounters challenges in safety‑critical applications, as performance still degrades under significant feature shifts. Additionally, the approach’s heavy reliance on domain expert reasoning, which is not always readily available, coupled with the high financial costs associated with fine‑tuning commercial LLM APIs, presents further practical constraints. Also, the nature of the data generation process based on the different energy rules in terms of the unknown anomalies is not clear. Quincozes \textit{et al.}~\cite{quincozes2023ereno} proposed the Efficacious Reproducer Engine for Network Operations (ERENO), a framework for generating realistic intrusion detection datasets specialized to IEC61850‑based standards. Their system synthesized traffic features by integrating data from both network and physical domains, thereby facilitating cross‑protocol detection between GOOSE and SV messages. The framework successfully generated datasets that model eight distinct use cases, including common attack types as well as normal network traffic and demonstrates that the integration of enriched substation features can enhance intrusion detection performance. Nevertheless, their implementation faces gaps in effectively detecting sophisticated masquerade attacks that are engineered to mimic legitimate behavior, and the current proof‑of‑concept primarily addresses attack scenarios based on illegitimate GOOSE messages, leaving other potential attack vectors and protocols less explored.
\subsection{Contributions}
According to the current research gaps, the enhancements of this research can be classified based on the data generation and efficiency of the proposed GenAI-based ADS based on the ToD framework~\cite{zaboli2025advanced}. Therefore, the main strengths of the proposed methodology using the previously proposed GenAI-based ADS can be summarized as follows:
\begin{itemize}
    \item \textbf{An AATM technique for synthetic data generation:} A novel perturbation and mutation-based synthetic data generation methodology is proposed to address the critical challenge of insufficient and imbalanced IEC61850-based communication datasets in digital substations. The AATM technique employs protocol-aware transformation functions that generate realistic zero-day attack patterns while maintaining strict adherence to GOOSE protocol rules through gradient-guided perturbations and categorical feature mutations. This method can outperform existing approaches including conditional generative adversarial network (CGAN). This methodology establishes a robust foundation for generating protocol-compliant synthetic datasets essential for training advanced ADSs against evolving cyber threats in digital substations.
    \item \textbf{A validation of the GenAI-Based ADS considering the ToD framework, compared with ML-based models through standard and advanced performance metrics:} This method advances traditional ML-based ADSs by utilizing the contextual comprehension abilities of GenAI, along with integrating expertise from domain specialists through the continuous learning. Hence, a comparative analysis of the GenAI-based ADS with a ToD configuration is conducted to demonstrate the superiority of rule-based GenAI frameworks over traditional ML algorithms including Feedforward Neural Networks (FNN), Recurrent Neural Networks (RNN), and Support Vector Machines (SVM) for an AD process in IEC61850 communications. In this case, the AATM generated GOOSE datasets are considered for the comparative analysis which have better balance rate (BR) and realism rate (RR). The evaluation employs both standard and advanced metrics to provide comprehensive assessment of detection capabilities. This validation establishes that incorporating domain-specific rules and contextual understanding through GenAI significantly enhances AD performance beyond what traditional ML algorithms can achieve.
\end{itemize}
\subsection{Paper Organization}
The rest of this paper is organized as follows: Section~\ref{section-AATM} presents the IEC61850-based multicast data generation techniques, particularly GOOSE messages. Further, this section provides different steps of the generation process based on the proposed and current techniques, GOOSE rules, and validation framework. Section~\ref{section-genai-based-ads} demonstrates the description of the proposed GenAI-based ToD framework and modeling of this structure alongside ML-based ADSs. The results and discussion of the proposed AATM data generation techniques in terms of the balance and realistic aspects are given in Section~\ref{section-results}. Also, a comparison of the proposed GenAI- and ML-based ADSs is implemented in this section considering the novel GOOSE message generation technique. Finally, the concluding remarks, along with potential directions for future research, are presented in Section~\ref{conclusion-section}.
\section{Proposed IEC61850-based Multicast Data Generation Technique in Digital Substations} \label{section-AATM}
In a controlled and realistic environment, a HIL testbed is crucial for evaluating the interaction between cyber attacks and the robustness of power systems. This real-time HIL testbed comprises an integration of various elements, such as hardware, software, communication protocols, and simulation technologies, all integrated with GPS synchronization. The incorporation of these elements is essential for exploring the real-time dynamics inherent in communication and information processing. This understanding is critical for the analysis of cyber attacks, enhancing detection protocols, and developing robust strategies for effective mitigation~\cite{hong2021implementation}. The configuration of the HIL testbed comprises diverse elements such as protection IEDs, SDN switches, a GPS unit, a merging unit IED, a SCADA system, a real-time digital simulator, and an amplifier. The system utilizes a SCADA-based distribution management system (DMS) that gathers measurements and implements control commands utilizing DNP3 protocols. The designed IEDs possess the function of dispatching control signals to circuit breakers (CBs). A CB, in turn, is engineered to react to GOOSE messages by communicating its operational status—either open or closed—back to the protective IEDs. Moreover, the merging unit IED is responsible for transmitting digital current and voltage measurements from the digital real-time simulator to the protective IEDs by employing the amplifier~\cite{choi2020multi, ten2011anomaly}. It is important to mention that specifics about the HIL testbed are outside the scope of this study, as this research primarily focuses on techniques for data pre-processing and generation.
Within the HIL test environment, Wireshark (a tool for analyzing network packets) enables a comprehensive capture of communication packets. This procedure entails real-time observation and examination of network traffic in the HIL testbed, facilitating comprehensive tracking and recording of packet flows. Utilizing the functionalities of Wireshark, researchers can capture a snapshot of communication flows, facilitating a deeper comprehension of the interactions among various components within the test environment. This systematic approach guarantees precise extraction of essential packets, thereby enriching the research endeavor with significant insights into the operational behaviors of cyber-physical systems (CPSs)~\cite{zaboli2024chatgpt}. Detailed information on the datasets, the procedures for extraction, and the outlined features, along with the GOOSE rules for both normal and abnormal patterns, will be discussed in the forthcoming section. This methodology can similarly be applied to other multicast messages (e.g., SV messages) due to their unique rules and characteristics.
Hence, this section introduces an innovative analytical framework for an analysis of AD processes in IEC61850-based communication messages (specifically GOOSE messages). An in-depth analysis is outlined as follows:
\subsection{Synthetic Balanced and Realistic Data Generation Process}
\begin{figure}[!h]
\centerline{\includegraphics[width=0.7\columnwidth]{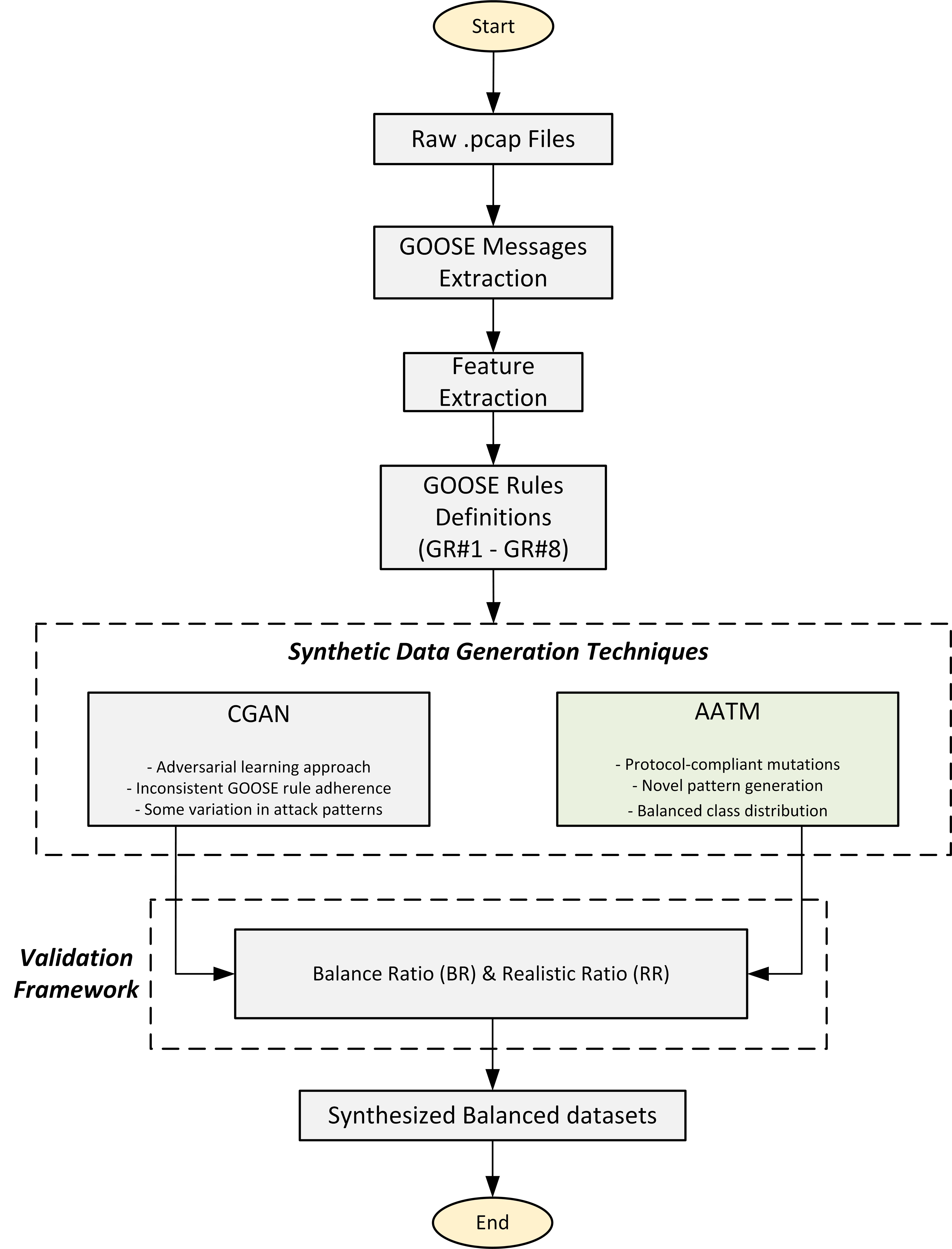}}
\caption{A systematic workflow of the proposed methodology for generating balanced and realistic GOOSE datasets.}
\label{GOOSE_preprocess_validation}
\end{figure}

\begin{figure}[!h]
\centerline{\includegraphics[width=1.0\columnwidth]{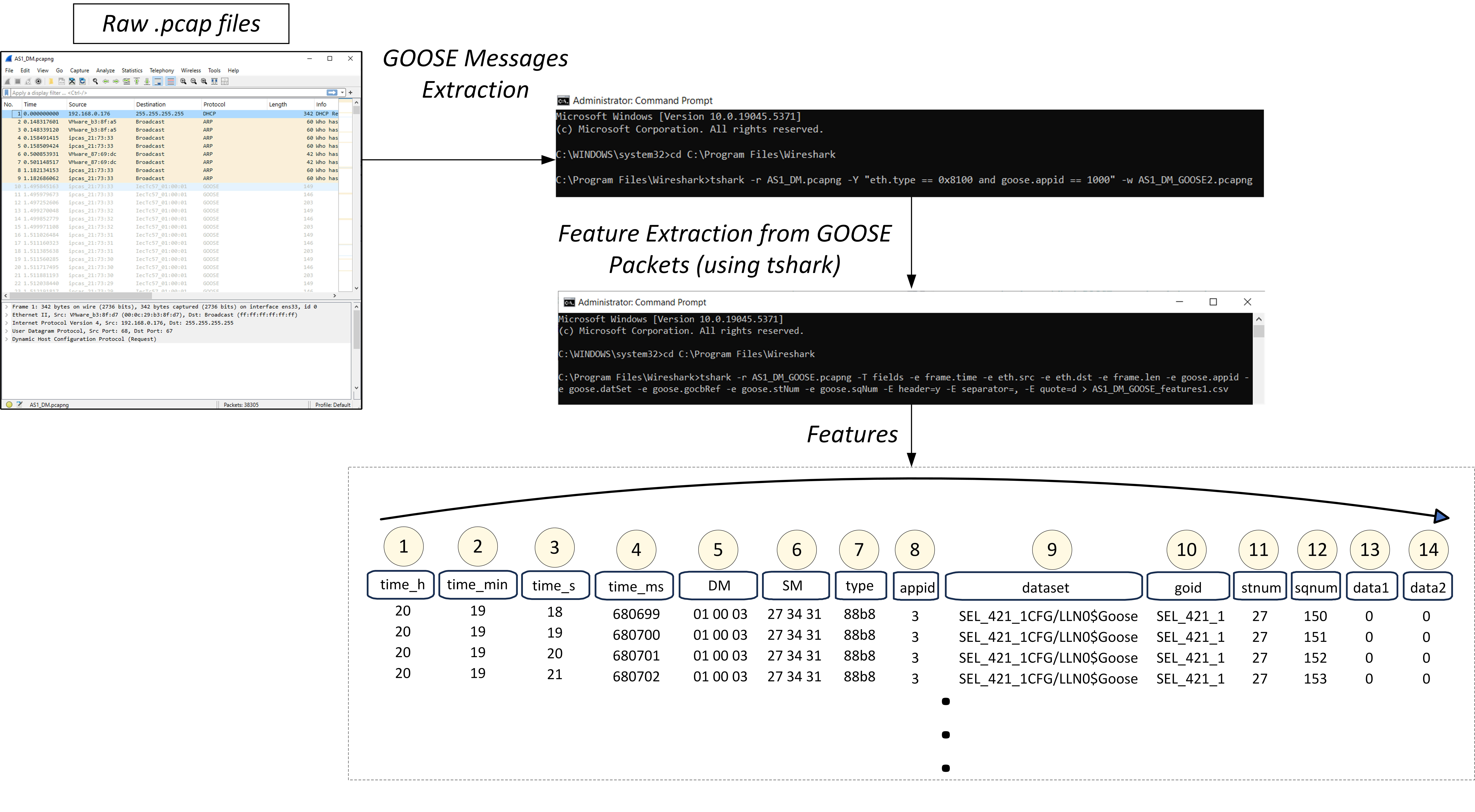}}
\caption{The process of GOOSE packets and features extraction using tshark.}
\label{tshark}
\end{figure}

This part shows the process for the generation of the GOOSE datasets in a way that can meet the requirements for the generation of realistic zero-day attacks and balancing issues. Hence, a comparison of the proposed technique known as the AATM with another methodology (i.e., CGAN~\cite{lim2024future, yuan2023data}) is carried out to show the better performance of this proposed method. Hence, a general framework of the data generation part considering the contributions and different steps is illustrated in Fig.~\ref{GOOSE_preprocess_validation}.
This pipeline begins with a raw packet capture step, proceeds through feature extraction and GOOSE rules definitions. Before the GOOSE rules definitions, it is necessary to provide details of GOOSE features in a sample dataset. The GOOSE packet data encompasses $14$ distinct data types, designated by the features extracted using tshark (a terminal-oriented version of Wireshark) which is shown in Fig.~\ref{tshark}~\cite{zaboli2024chatgpt}.
The temporal attribute meticulously logs the exact transmission instance of a packet, detailing the time in hours, minutes, seconds, and milliseconds to ensure comprehensive precision. The abbreviations DM and SM stand for the destination and source media access control (MAC) addresses, respectively, acting as essential identifiers in the communication framework. In particular, the designated DM address associated with GOOSE messages is referred to as ($01 00 03$), and is directed at devices associated with this specific MAC address, whereas the SM address is depicted as $27 34 31$, which determines the transmitting IED. In GOOSE packet classification, the type indicator is specified by the value $88b8$. Furthermore, for GOOSE communications, the Application identifier (APPID) is set to $3$. The dataset name and GOOSE identification are expressed by the dataset and goid attributes, which depend on the DM address. Additionally, state number (stNum) and sequence number (sqNum) are utilized in the context of GOOSE messages. In the analysis phase, data1 ($d1$) and data2 ($d2$) are considered, both of which are derived as binary features from GOOSE packets.
Given a ``\textit{.pcap}'' file, $P$ containing all packets in which $p$ represents only each individual packet. The GOOSE set is defined as $G$ that only includes packets that meet specific criteria as described in Eq.~(\ref{tshark1})~\cite{biswas2019synthesized, hong2014integrated}:
\begin{equation} \label{tshark1}
G = \{p \in P \mid p.\text{eth\_type} = 0\times8100 \land p.\text{appid} = 1000\}
\end{equation}
The first condition shows the Ethernet type that must be $0 \times 8100$ for GOOSE messages, and the second condition is relevant to GOOSE APPID which should $1000$ according to messages. By meeting these two conditions, the system guarantees extracting only GOOSE packets. This process can be carried out using tshark, enabling the capture of packet data from live networks as well as the reading of packets from pre-existing capture files. It provides the option to output a decoded representation of the data to the standard output or to write the packets to a file for subsequent analysis. Further, it can extract the features of data as well as a data conversion to other formats (e.g., .csv format). This extraction process ensures the capture of all relevant GOOSE messages while filtering out other traffic types. 
In the subsequent phase, GOOSE guidelines (\textit{GR\#1} through \textit{GR\#8}) are outlined to address a range of both normal and abnormal scenarios. Eqs.~(\ref{GR1})--(\ref{GR8}) demonstrate the GOOSE guidelines utilized in this study to examine the various anomalies within datasets as follows~\cite{zaboli2024leveraging, zaboli2025advanced}:

\textit{GR\#1}: In the event that sequential data packets possess the same \textit{DM} and \textit{SM} characteristics, the \textit{sqNum} parameter must be incremented. If there are discrepancies, it is indicative of an abnormality.
\begin{equation} \label{GR1}
\small
GR\#1(G_i, G_{i-1}) = 
\begin{cases}
    1, & \text{if } DM_i = DM_{i-1} \land  SM_i = SM_{i-1} \land  sqNum_i = sqNum_{i-1} + 1 \\
    0, & \text{otherwise}
\end{cases}
\end{equation}

\textit{GR\#2}: In GOOSE communications, an anomaly that suggests a DI attack occurs when there is a transition in a data value, such as \textit{data1} (\textit{d1}) or \textit{data2} (\textit{d2}), from $0$ to $1$ or from $1$ to $0$, while the \textit{stNum} remains unchanged and the \textit{sqNum} is consistently incremented in sequence.
\begin{equation} \label{GR2}
\small
GR\#2(G_i, G_{i-1}) =
\begin{cases}
1, & \text{if } (d1_i \neq d1_{i-1} \vee d2_i \neq d2_{i-1}) \wedge (stNum_i = stNum_{i-1}) \wedge 
 (sqNum_i = sqNum_{i-1} + 1) \\
0, & \text{otherwise}
\end{cases}
\end{equation}

\textit{GR\#3}:
Within GOOSE messages where the DM and SM are identical, the \textit{stNum} is expected to either stay constant under normal scenarios or increment when data changes occur. A reduction in \textit{stNum} is considered irregular unless it results from a system rollover—specifically, when \textit{stNum} attains its maximum value of $2^{32} -1$, it resets to $0$ in the next GOOSE transmission—or from a valid system reboot.
\begin{equation} \label{GR3}
\small
GR\#3(G_i, G_{1:i-1}) =
\begin{cases}
1, & \text{if } DM_i = DM_{i-1} = ...= DM_{i-n} \& SM_i = SM_{i-1} = SM_{i-n} \land \\ & (stNum_i = stNum_{i-1} \lor stNum_i = stNum_{i-1} + 1 \lor stNum_i = 0 \land \\
  & stNum_{i-1} = 2^{32}-1)) \\
0, & \text{otherwise}
\end{cases}
\end{equation}
\textit{GR\#4}: Variations in the parameters \textit{DM}, \textit{SM}, \textit{type}, \textit{appid}, \textit{dataset}, or \textit{goid} imply the occurrence of abnormal conditions.
\begin{equation} \label{GR4}
\small
GR\#4(G_i, G_{i-1}) = 
\begin{cases}
    1, & \text{if } DM_i = DM_{i-1} \land SM_i = SM_{i-1} \land type_i = type_{i-1} \land dataset_i = dataset_{i-1} \\
    0, & \text{otherwise}
\end{cases}
\end{equation}

\textit{GR\#5}: Although GOOSE messages transmit timestamps in a binary format, it is essential that the timestamps extracted and subsequently decoded within this dataset adhere to a uniform presentation format, specifically reflecting hours, minutes, seconds, and milliseconds (e.g., HH:MM:SS.mmm). 
\begin{equation} \label{GR5}
\small
    GR\#5(time_i) = 
    \begin{cases}
        1, & \text{if } time_i \text{ is in format} \\
        0, & \text{otherwise}
    \end{cases}
\end{equation}

\textit{GR\#6}: Although it is recognized that the frequent occurrence of GOOSE messages is typical during protection operations—such as those involving busbar protection with several breaker activations lasting up to $500$ ms—this framework analyzes the patterns in message frequency. The core of this monitoring focuses on the count of consecutive messages, identified by their timestamps at the millisecond level, and sets a standard threshold of $10$ instances occurring within a $10$ $\mu$s window. 
\begin{equation} \label{GR6}
\small
    GR\#6(G_{i-9:i}) = 
    \begin{cases}
        1, & \text{if } \forall j \in [i-9, i-1]: time_{j+1} - time_j \leq 10\mu s \\
        0, & \text{otherwise}
    \end{cases}
\end{equation}

\textit{GR\#7}: A data transmission interruption extending beyond $10$ seconds serves as a sign of an anomalous condition.
\begin{equation} \label{GR7}
\small
    GR\#7(G_i, G_{i-1}) = 
    \begin{cases}
        1, & \text{if } time_i - time_{i-1} \leq 10s \\
        0, & \text{otherwise}
    \end{cases}
\end{equation}

\textit{GR\#8}: In the context of GOOSE messages, a transition in a data value (e.g., \textit{data1} or \textit{data2}) from $1$ to $0$ or from $0$ to $1$, coupled with an unchanged \textit{stNum} and a \textit{sqNum} that remains at $0$, signifies an irregularity suggestive of a possible RE attack.
\begin{equation} \label{GR8}
\small
GR\#8(G_i, G_{i-1}) =
\begin{cases}
1, & \text{if } (d1_i \neq d1_{i-1} \vee d2_i \neq d2_{i-1}) \wedge (stNum_i = stNum_{i-1}) \wedge (sqNum_i = sqNum_{i-1}) \\
0, & \text{otherwise}
\end{cases}
\end{equation}
According to Fig.~\ref{GOOSE_preprocess_validation}, the next step presents two different techniques for synthetic data generation including CGAN and the proposed technique known as AATM. Compared with the CGAN technique, the AATM technique is characterized by its use of transformation functions instead of NNs, eliminating the need to train a model from scratch while operating through directed perturbations of existing samples under rule-based constraints and utilizing gradient information (i.e., derivatives or rates of change of a function) to guide these perturbations without building or training an NN architecture. Also, there are other techniques (e.g., multi-mixing~\cite{manzoor2024zero}) which cannot meet requirements for balance and realistic concerns because the nature of this method is based on simple combinations with some weighting coefficients. While the multi-mixing technique merely interpolates between existing samples and cannot explore beyond known patterns, AATM applies protocol-aware transformations that can generate novel attack vectors while maintaining protocol compliance. Then, the generated datasets undergo evaluation for balance and realism checks before being used in the ADS frameworks. The following subsections represent the mathematical modeling of the CGAN and proposed AATM techniques along with GOOSE guidelines, considering the BR and RR functions to check the validity of synthesized datasets.

\subsubsection{\textbf{CGAN Technique}}
The CGAN formulation for GOOSE message generation employs an adversarial approach, as shown in Eq.~(\ref{CGAN_1})~\cite{yuan2023data, sauber2022use}:

\begin{equation} \label{CGAN_1}
\begin{split}
\min_G \max_D V(D,G) = \mathbb{E}_{x \sim p_{data}(x)}[\log D(x|y,c)] + \mathbb{E}_{z \sim p_z(z)}[\log(1 - D(G(z|y,c)))]
\end{split}
\end{equation}

Where:
\begin{itemize}
    \item $z \sim \mathcal{N}(0, I_d)$: A \textit{d}-dimensional random noise vector sampled from a standard normal distribution, serving as the randomization source for generating diverse GOOSE messages while maintaining desired characteristics. Each dimension influencing different aspects of the synthetic message.
    \item $y \in \{0,1\}^m$ is a vector representing the attack class (e.g., DI, RE attack, and DoS) or normal traffic, allowing the CGAN to generate class-conditional samples. For GOOSE messages, $m$ typically ranges from $5-8$ depending on how many attack types are modeled.
   \item $c \in \{0,1\}^p$ is a binary vector encoding the GOOSE context. These contextual factors ensure generated messages reflect realistic operational scenarios within IEC61850 environments.
   \item $G: \mathbb{R}^d \times \{0,1\}^m \times \{0,1\}^p \rightarrow \mathbb{R}^{14}$: The generator function that transforms the noise vector, conditioned on attack class and protocol context, into a synthetic 14-dimensional GOOSE message containing all required features.
   \item $D: \mathbb{R}^{14} \times \{0,1\}^m \times \{0,1\}^p \rightarrow [0,1] \times \{0,1\}^8 $: The discriminator function that evaluates whether a given GOOSE message appears realistic given its claimed attack class and protocol context; outputting a probability between $0$ and $1$ considering the GOOSE rules where higher values indicate the message appears genuine rather than synthetic.
\end{itemize}
The adversarial loss for statistical realism and the rule compliance loss for protocol validity constitute the total loss function as Eq.~(\ref{L_total_CGAN}):

\begin{equation} \label{L_total_CGAN}
\mathcal{L}_{total} = \mathcal{L}_{adv} + \lambda_{rules}\mathcal{L}_{rules}
\end{equation}

Where the rule compliance loss has different weights based on the security importance, as shown in Eq.~(\ref{rule compliance loss}):

\begin{equation} \label{rule compliance loss}
\mathcal{L}_{rules} = \sum_{i=1}^8 w_i(1 - GR_i(G(z|y,c)))^2
\end{equation}

The adversarial loss ensures generated messages match the statistical properties of real GOOSE traffic, while the rule compliance loss (weighted at $\lambda_{rules}$ = $8.0$) enforces adherence to the eight GOOSE rules. The generator is directly penalized for producing non-compliant GOOSE messages via the weighted sum of squared errors. Conversely, the discriminator learns about rule compliance indirectly by comparing real rule-following samples to generated ones.
Within the rule compliance term, individual weights (i.e., $0 < \omega_i < 2$) reflect each rule's security importance: critical integrity rules (GR\#3, GR\#4) receive the highest weights ($2.0$), RE and data manipulation rules (GR\#2, GR\#8) receive moderate-high weights ($1.5-1.8$), and formatting/timing rules receive lower weights ($0.8-1.0$). This formulation creates a balanced objective that prioritizes both realistic message generation and protocol compliance, with emphasis on security-critical constraints.

\subsubsection{\textbf{Proposed AATM Technique}}
This proposed approach employs a protocol-aware transformation function which is particularly useful for communication messages. As communication messages have a specific pattern with some subtle changes, a proper definition of perturbations can help to distinguish these small changes. These subtle changes can even happen in some numbers and/or letters in the different features (both numerical and categorical) which the CGAN technique might fail to understand; these variations, specifically in cases of the generation of realistic zero-day attacks and a high similarity between the normal datasets and some types of attacks (e.g., RE attacks)~\cite{bhuyan2013network, garcia2014empirical}. The following information shows the process of different steps in the proposed method.
\paragraph{Vector Representation
} Suppose a GOOSE dataset is represented as:

\begin{equation}
x_i = [x_i^{num}, x_i^{cat}]
\end{equation}

Where $x_i^{num}$ represents 9 numerical features (e.g., time, appid, and stNum) and $x_i^{cat}$ represents 5 categorical features (e.g., DM, SM, and type). In the numerical part, three different functions including protocol compliance, balance, and novelty functions are presented in Eqs.~(\ref{protocol_AATM})-~(\ref{novelty_AATM}) as follows:

\begin{equation} \label{protocol_AATM}
f_{protocol}(x_i) = \sum_{j=1}^{m} w_j \cdot GR_j(x_i)
\end{equation}

\begin{equation} \label{balance_AATM}
f_{balance}(x_i) = -\sum_{c \in C}\left|p_c - \frac{1}{|C|}\right|
\end{equation}

\begin{equation} \label{novelty_AATM}
f_{novel}(x_i) = \min_{x_j \in X}d(x_i, x_j)
\end{equation}

The protocol compliance function, $f_{protocol}^{num}(x_i)$, uses weighted rule compliance metrics $GR_j(x_i)$ to guide perturbations toward protocol-valid regions, ensuring attacks remain credible by preserving necessary relationships between features. The balance function, $f_{balance}(x_i)$, calculates the negative sum of deviations between current class proportions, $p_c$, and ideal uniform distribution, $\frac{1}{|C|}$, directing mutations toward minority attack classes to create diverse datasets with equal representation across attack types; and the novelty function, $f_{novel}(x_i)$, calculates the minimum distance between a candidate sample and all training examples, encouraging exploration of unexplored feature space regions to discover attack vectors not present in existing datasets, thereby enhancing the model's ability to generate previously unknown attack patterns that could evade traditional detection systems while maintaining structural validity. The selection of weights, $w_j$, for the eight GOOSE protocol rules is crucial for effective attack generation and follows a priority-based approach: higher weights ($0.15-0.2$) are assigned to critical sequence rules (i.e., stNum \& sqNum) and timestamp validation as these are fundamental to the protocol's integrity and frequently monitored by ADSs. The medium weights (0.1) are given to APPID compliance and MAC address validity as they represent network-level identifiers that must appear legitimate. This weighted approach allows the proposed framework to generate attacks that target specific vulnerabilities while maintaining sufficient protocol compliance to avoid simple detection, creating attack effectiveness.

\paragraph{Gradient Computation}
The base gradient for perturbation in the numerical part can be mentioned as Eq.~(\ref{gradient_num}) which states a combination of presented functions along with different hyperparameters.

\begin{equation} \label{gradient_num}
\begin{split}
\delta_{base}^{num}(x_i) = \alpha \cdot \nabla_{x}f_{protocol}(x_i) + \beta \cdot \nabla_{x}f_{balance}(x_i) + \gamma \cdot \nabla_{x}f_{novel}(x_i)
\end{split}
\end{equation}

The hyperparameters $\alpha$, $\beta$, and $\gamma$ in this equation are typically set to $\alpha=0.4$, $\beta=0.3$, and $\gamma=0.3$, respectively. This weighting prioritizes protocol validity ($\alpha=0.4$) to ensure attacks remain credible within the IEC61850 framework, while equally distributing the remaining influence between class balance and novelty exploration ($\beta=\gamma=0.3$). These values can be adjusted based on specific attack objectives. For example, an increased $\beta$ is required when greater attack diversity is needed, or raising $\gamma$ is needed when novel attack discovery is prioritized over keeping the protocol compliance. The zero-day perturbation formulation represents the core innovation of the AATM approach, as stated in Eq.~(\ref{Zero-day_perturbation}).

\begin{equation} \label{Zero-day_perturbation}
\delta_{zero}^{num}(x_i) = \delta_{base}^{num}(x_i) - \lambda \cdot \sum_{r \in R_{target}} \nabla_{x}GR_r(x_i)
\end{equation}

According to this, the base perturbation is modified by subtracting the weighted gradient of targeted protocol rules, with $\lambda$ controlling the violation strength and $R_{target}$ specifying which rules to deliberately violate; this approach effectively pushes samples away from compliance with selected rules while maintaining overall protocol validity. The new numerical features are then generated using the projection function (Eq.~(\ref{projection_num})) which applies the zero-day perturbation to the original numerical features and projects the result into the valid numerical feature space through $P^{num}$.

\begin{equation} \label{projection_num}
x_{new}^{num} = P^{num}(x_i^{num} + \delta_{zero}^{num}(x_i))
\end{equation}

This ensures that perturbed values maintain protocol compliance except for the deliberately targeted violations, producing realistic attack vectors that specifically exploit the targeted protocol vulnerabilities.

\paragraph{Categorical Feature Processing}
A categorical transition matrix, $T_{r,j}$ to show how feature $j$ should change to affect rule $r$ as Eq.~(\ref{Transition_cat}):

\begin{equation} \label{Transition_cat}
T_{r,j} = \begin{bmatrix} 
T_{1,1} & T_{1,2} & \cdots & T_{1,5} \\
T_{2,1} & T_{2,2} & \cdots & T_{2,5} \\
\vdots & \vdots & \ddots & \vdots \\
T_{m,1} & T_{m,2} & \cdots & T_{m,5}
\end{bmatrix}
\end{equation}

Next, the categorical mutations can be defined as Eq.~(\ref{cat_mutations}) to generate zero-day attacks in terms of non-numerical parts. 

\begin{equation} \label{cat_mutations}
M_{zero}(x_i) = \begin{bmatrix} m_1 \\ m_2 \\ \vdots \\ m_5 \end{bmatrix}
\end{equation}

Where each $m_j$ is calculated based on Eq.~(\ref{m_j_cat}), and the base mutation combines protocol, balance, and novelty objectives:

\begin{equation} \label{m_j_cat}
m_j = \text{base\_mutation}_j - \lambda \cdot \sum_{r \in R_{target}} T_{r,j}
\end{equation}

\begin{equation} \label{base_mutation_cat}
\begin{split}
\text{base\_mutation}_j = \alpha \cdot \text{protocol\_mut}_j + \beta \cdot \text{balance\_mut}_j + \gamma \cdot \text{novelty\_mut}_j
\end{split}
\end{equation}

Now, a new categorical feature vector can be given according to Eq.~(\ref{cat_new_features}) that describes how categorical features in GOOSE messages are mutated during zero-day attack generation. It works by first finding the index position of the current categorical values, applying the calculated mutation value, and then selecting the corresponding categorical value at the new index position while ensuring it remains within the valid set of possible values through the modulo operation.

\begin{equation} \label{cat_new_features}
x_{new,j}^{cat} = C_j[(I_j(x_{i,j}^{cat}) + \text{Round}(m_j)) \mod |C_j|]_{1 \times 2}
\end{equation}

Where $C_j$ is the set of valid values for the categorical feature $j$, and $I_j$ maps the categorical value to its index. Further, ``Round'' converts the mutation to an integer value. Also, the modulo operation helps to ensure that the resulting index does not exceed the bounds of the array. Suppose that multiple values in the DM column are as follows:
\begin{mdframed}
$C_{DM} = \text{[``01 00 03,'' ``01 00 04,'' ``01 00 05'']}$ with $|C_{DM}| = 3$. 

- For a row with DM = ``01 00 03'' and $m_{DM} = 1.5$:

- $I_{DM}$(``$01 00 03$'') $= 0$

- $Round (m_{DM}) = 2$

- New index $= 0+2=2$

- $x_{new,DM} = C_{DM}[2] =$ ``$01 00 05$''

However, if $m_{DM}$ $= 2.5$:

- Round($m_{DM}$) $= 3$

- New index $= 0 + 3 = 3$
\end{mdframed}
This would be an error since the valid indices are only $0$, $1$, and $2$. This is why the modulo operation is important to bound the indices to the available size. Hence, the complete process of the proposed technique can be summarized as follows to clearly show different steps in plain language.
\begin{enumerate}
    \item Split the original GOOSE message into numerical and categorical components
    \item Compute numerical perturbation $\delta_{zero}^{num}(x_i)$
    \item Compute categorical mutations $M_{zero}(x_i)$
    \item Apply perturbation to numerical features
    \item Apply mutations to categorical features
    \item Combine into final vector $x_{new} = [x_{new}^{num}, x_{new}^{cat}]$
\end{enumerate}

This process demonstrates the AATM's ability to precisely target specific protocol rules for violation while maintaining overall message validity, creating attacks that are challenging to detect. The next part shows the validation process including the BR and RR for further processing in the first layer of this framework.
\subsubsection{Validation Framework}
A validation framework focusing exclusively on two critical quality metrics, including BR and RR, is presented to evaluate the quality of generated synthesized datasets. 

\textbf{\textit{- Balance Rate Assessment of Generated Datasets}}

The BR quantifies dataset balance using Eq.~(\ref{balance_rate_shannon}) that the first segment focuses on the relationship between the most extreme classes in the dataset, and the second segment shows the normalized Shannon entropy~\cite{buda2018systematic, yu2022pricing}.

\begin{equation} \label{balance_rate_shannon}
BR(X) = \frac{1}{2} \left( \frac{\min_i(n_i)}{\max_i(n_i)} + E \right)
\end{equation}

Where $n_i$ is the number of samples in class $i$, $\frac{\min_i(n_i)}{\max_i(n_i)}$ is the inverse of the imbalance ratio (ranges from 0 to 1), $E = -\frac{\sum_{i=1}^K p_i \log_2(p_i)}{\log_2(K)}$ is the normalized Shannon entropy, $p_i = \frac{n_i}{\sum_{j=1}^K n_j}$ is the proportion of samples in class $i$ and $K$ is the total number of classes. The first part directly measures how much smaller your least represented class is compared to your most common class. This is particularly important for GOOSE message security applications because it highlights if any attack class is significantly infrequent compared to normal traffic. While the first part only looks at the extremes, Shannon entropy examines the distribution across all classes simultaneously. It measures how evenly distributed the samples are across every class, not just the smallest and largest ones. Also, $BR(X)$ has a range of $0 < BR(X) \leq 1$ which $0$ and $1$ show the completely imbalanced and perfectly balanced datasets, respectively. Hence, this combination shows two key aspects of class balance in terms of the ratio between least and most frequent classes, addressing extreme imbalances as well as the overall diversity of the distribution, capturing how evenly samples are distributed across all classes.

\textbf{\textit{- Realism Rate Assessment of Generated Datasets}}

For synthesized GOOSE datasets, an RR can be presented that evaluates protocol compliance through essential rule verification as shown in Eq.~(\ref{realism_rate}) where $GR_i(x)$ evaluates compliance with each essential GOOSE protocol rule. 

\begin{equation} \label{realism_rate}
RR(x) = \prod_{i=1}^8 GR_i(x)
\end{equation}

Moreover, Eq.~(\ref{GR_realism}) expresses the scoring mechanism for each protocol rule using function $GR_i(x)$, which assigns a value of $1$ when rule $i$ is completely satisfied, thus signifying perfect compliance, while an exponential penalty is applied in cases of rule violation. 

\begin{equation} \label{GR_realism}
GR_i(x) = 
\begin{cases}
1, & \text{if rule $i$ is fully satisfied} \\
e^{-\lambda_i \cdot V_i(x)}, & \text{Otherwise}
\end{cases}
\end{equation}

$V_i(x)$ represents the normalized severity of the violation (scaled between $0$ and $1$) and $\lambda_i$ is the importance weight assigned to that specific rule. This formulation effectively creates a continuous scoring mechanism that severely penalizes violations of critical rules (those with higher $\lambda_i$ values) while allowing minor deviations in less critical aspects, making it particularly suitable for evaluating synthesized GOOSE messages~\cite{yang2015cybersecurity}.

\begin{equation*}
\lambda_i = 
\begin{cases}
5, & \text{for critical integrity rules: } GR_3, GR_8 \\
3, & \text{for important rules: } GR_1, GR_2, GR_4 \\
2, & \text{for structural rules: } GR_5, GR_7 \\
1, & \text{for timing pattern rules: } GR_6
\end{cases}
\end{equation*}

The violation severity, $V_i(x)$, provides a standardized illustration for quantifying deviations from GOOSE protocol rules, with each measure ($V_1(x)-V_8(x)$) targeting a specific aspect of message integrity as shown in Table~\ref{tab:violation}. $V_1(x)$ assesses sqNum correctness, $V_2(x)$ detects DI patterns, $V_3(x)$ monitors stNum consistency, $V_4(x)$ measures unexpected field changes as a proportion, $V_5(x)$ validates timestamp formatting, $V_6(x)$ evaluates message frequency anomalies on a continuous scale, $V_7(x)$ quantifies communication gaps relative to acceptable thresholds, and $V_8(x)$ identifies RE attack signatures.
\begin{table}[!h]
\small
\caption{The violation severity measures for GOOSE protocol rules.}
\label{tab:violation}
\centering
\begin{tabular}{|c|c|c|}
\hline
\bfseries Measure & \bfseries Description & \bfseries Value \\
\hline
$V_1(x)$ & sqNum increment violation & \makecell{Binary: 1 if sqNum does not increment correctly,\\ 0 otherwise.} \\
\hline
$V_2(x)$ & \makecell{Data change with unchanged\\ stNum violation} & \makecell{Binary: 1 if data change logic is violated,\\ 0 otherwise.} \\
\hline
$V_3(x)$ & stNum monotonicity violation & \makecell{Binary: 1 if stNum decreases inappropriately,\\ 0 otherwise.} \\
\hline
$V_4(x)$ & Field integrity violation & $\frac{(\text{Number of unexpectedly changed critical features})}{(\text{Total number of critical features})}$ \\
\hline
$V_5(x)$ & Timestamp format violation & \makecell{Binary: 1 if timestamp format is invalid,\\ 0 otherwise.} \\
\hline
$V_6(x)$ & Message frequency violation & $\min(1, \frac{\text{Observed frequency}}{\text{Maximum threshold}} - 1)$ \\
\hline
$V_7(x)$ & Temporal gap violation & \makecell{$\min(1, \frac{\text{Gap duration} - 10s}{30s})$ for gaps $> 10$s} \\
\hline
$V_8(x)$ & RE attack indicator & \makecell{Binary: 1 if data changes without stNum \\increment and sqNum reset, 0 otherwise} \\
\hline
\end{tabular}
\end{table}
In this context, $X$ represents a single dataset (a group of messages), and $|X|$ is the number of messages within that specific dataset. Hence, the aggregate realism can be given as Eq.~(\ref{aggregated_realism}):

\begin{equation} \label{aggregated_realism}
RR(X) = \frac{1}{|X|}\sum_{x \in X} RR(x)
\end{equation}

In which $RR(X) \geq 0.95$ shows the excellent realism of the generated synthesized dataset. This methodology offers a robust and compelling strategy for assessing the authenticity of generated GOOSE datasets, all while ensuring adherence to fundamental protocol compliance standards is not compromised. To recap, this section presented the data pre-processing, a generation of normal and zero-day attacks, and data post-processing according to the balance and realism assessments. Different methods, including the proposed AATM technique, are represented according to the application for GOOSE messages in digital substations considering the rules and data features. The purpose of the upcoming sections is to use these pre-processed datasets for an AD based on the proposed frameworks.
\section{GenAI-based Anomaly Detection Systems in Digital Substations} \label{section-genai-based-ads}
ML-based ADSs have served as the basis for identifying anomalies within IEC61850 message frameworks. Although these approaches are known for their accuracy and reliance on data, they face a substantial limitation. Specifically, with the emergence of novel attack patterns (i.e., zero-day attacks), the models necessitate re-training. This requirement for model re-training is resource-intensive and time-consuming, introducing periods of vulnerability during which the system may not be equipped to cope with these unforeseen threats until they are integrated into the model’s knowledge repository~\cite{hong2022automated}.
Conversely, GenAI tools present a more flexible and versatile strategy. Differing from traditional ML models, GenAI systems possess the ability to comprehend context, thereby enabling them to identify and address emerging threats without the need for prior explicit training. This capability of contextual comprehension reduces the necessity for constant updates and retraining in the rapidly changing landscape of cyber threats. By interpreting and accommodating new data, GenAI tools offer a robust and effective approach to AD in digital substations, leveraging NLP capabilities~\cite{gill2023chatgpt}. According to the challenges mentioned and the comprehensive literature surveys, a GenAI-based ADS framework considering the ToD system is presented in~\cite{zaboli2025advanced} to show its performance evaluations.

Moreover, this section identifies critical sources for the dataset used in training the GenAI-enhanced ADS, along with the criteria applied for selecting these datasets. Compared to publicly accessible datasets, those sourced from the HIL testbed deliver high-resolution, authentic data that accurately represents actual substation operations~\cite{zaboli2024chatgpt}. However, it is challenging to have balanced and realistic datasets, without the existence of zero-day attacks. Hence, this paper evaluates the performance of the proposed ADS with 5,000 GOOSE datasets generated using the proposed AATM technique, which cover a wide range of classes including the normal and abnormal scenarios, and the BR and RR of the generated datasets are better than the CGAN approach.
\subsection{\textbf{GenAI-based Task-Oriented Dialogue ADS}}
\begin{figure}[!h]
\centerline{\includegraphics[width=1.0\columnwidth]{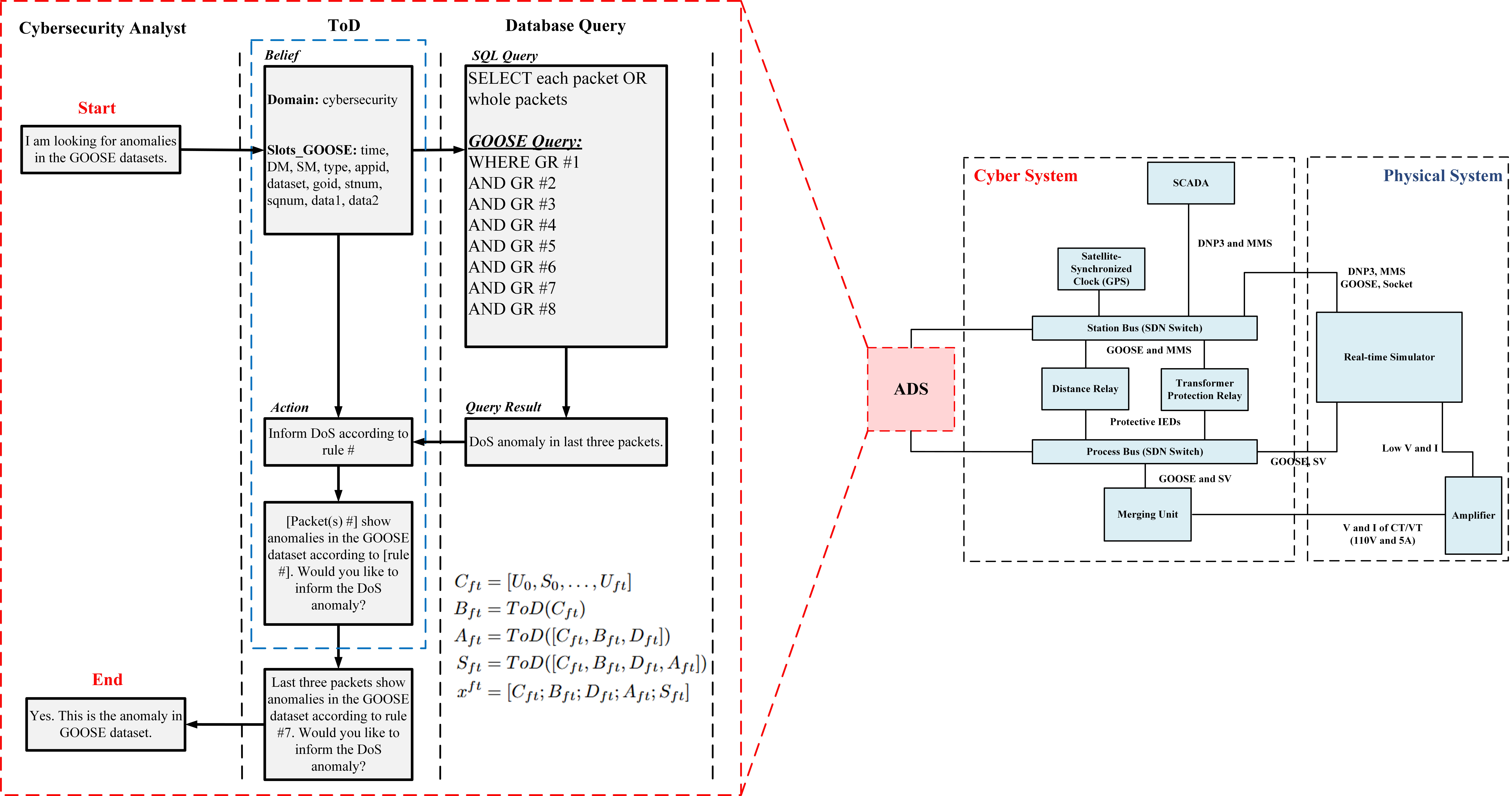}}
\caption{The proposed GenAI-based ToD ADS framework in the full training levels considering all GOOSE rules.}
\label{genai-tod-dissertation-ACCESS}
\end{figure}

The GenAI-powered ToD system's architectural framework incorporates advanced computational strategies through its hierarchical processing structure as proposed in our previous research~\cite{zaboli2025advanced}. The full training (FT) configuration of this framework is demonstrated in Fig.~\ref{genai-tod-dissertation-ACCESS} that considers all GOOSE rules. It is designed to optimally leverage the structured communication protocols prevalent in digital substation environments. 
This system's belief state is mathematically represented by aggregated sequences of packets and system states and is iteratively refined through rule-based SQL query executions that assess compliance to predefined operational constraints, including temporal synchronization, sqNum progression, and data integrity verification protocols\cite{zeng2024divtod, paek2024enhancing}. By integrating adaptive validation tools characterized by a hybrid function that includes belief states, dynamic conditions, and evolving rule sets, the system effectively distinguishes between legitimate operational behaviors and anomalies within the IEC61850 communication network. 

The framework efficiently interacts with CPS components through advanced interfacing techniques, which facilitate its integration with supervisory control systems. This enables real-time coordination of identified anomalies with responses from protective relays within the power grid setup. A continuous learning approach, utilizing iterative feedback loops that incorporate both detection outcomes and scenario simulations, progressively refines classification boundaries while preserving the computational efficiency necessary for urgent protection strategies. This detailed training approach guarantees that the system's query processing mechanism, which evaluates multi-conditional rules, attains an ideal balance between detection sensitivity and specificity. Therefore, it reinforces a strong security barrier for the protection of critical infrastructure, as indicated by performance metrics that surpass those of conventional detection methods. More information regarding the parameters and detailed analysis of different blocks is expanded in~\cite{zaboli2025advanced}. The intention of this paper is to evaluate the FT level of the GenAI-based ToD for the AD process, trained by the proposed AATM-generated GOOSE datasets, which are more realistic and balanced. Also, the previous research assessed the performance of different HITL and ToD in various GPT tools to show the better performance, in which the ToD framework implemented in Anthropic Claude Pro had the best performance in terms of the evaluation metrics as well as the accuracy. Hence, this paper uses the GenAI-based ToD framework implemented in Anthropic Claude Pro to make a comparison with other ML-based ADSs.
\subsection{ML-based ADSs vs. GenAI-based ADSs}
Traditional ML approaches, including FNNs, RNNs, and SVM, have shown promising results in detecting anomalies in IEC61850-based multicast messages. Hence, this part shows the modeling of these ML algorithms, particularly for GOOSE messages, as well as the general GenAI-based AD model. However, the emergence of GenAI technologies, particularly transformer-based architectures, has opened new possibilities for the AD concept. Hence, a general mathematical modeling of these ML algorithms, in addition to the GenAI-based system, is given below. More information and discussion about the results of the AD process considering different classes and their performance metrics are provided in the next section~\cite{su2024large}.
\subsubsection{Feedforward Neural Network (FNN)}
represents the fundamental architecture of DL, consisting of multiple layers of interconnected neurons where information flows one way from input to output without cycles or loops. In the context of the GOOSE AD process, FNNs are configured with an input layer that accepts normalized feature vectors extracted from GOOSE messages (e.g., timing parameters, sqNum, and data values), followed by multiple hidden layers with non-linear activation functions (typically ReLU), and an output layer with softmax activation for multi-class classification~\cite{huang2024design}. According to the application of the AD process in GOOSE messages, given an input vector $\mathbf{x} \in \mathbb{R}^n$, the FNN computes the output through successive layer transformations as Eq.~(\ref{fnn_basic}):

\begin{equation} \label{fnn_basic}
\mathbf{h}^{(l)} = f^{(l)}(\mathbf{W}^{(l)}\mathbf{h}^{(l-1)} + \mathbf{b}^{(l)})
\end{equation}

where $\mathbf{h}^{(l)}$ represents the activation of layer $l$, $\mathbf{W}^{(l)}$ is the weight matrix, $\mathbf{b}^{(l)}$ is the bias vector, and $f^{(l)}$ is the activation function. For the AD process, the network is trained to minimize the reconstruction error through the loss function in Eq.~(\ref{fnn_loss}):

\begin{equation} \label{fnn_loss}
\mathcal{L}_{FNN} = \frac{1}{N}\sum_{i=1}^{N} \|\mathbf{x}_i - \hat{\mathbf{x}}_i\|^2 + \lambda\sum_{l=1}^{L}\|\mathbf{W}^{(l)}\|_F^2
\end{equation}

where $\hat{\mathbf{x}}_i$ is the reconstructed input and $\lambda$ controls regularization, thus,the anomaly score can be computed as Eq.~(\ref{fnn_anomaly_score}):

\begin{equation} \label{fnn_anomaly_score}
A_{FNN}(\mathbf{x}) = \|\mathbf{x} - \hat{\mathbf{x}}\|^2
\end{equation}

\subsubsection{Recurrent Neural Network (RNN)} is designed to process sequential data by forming directed graphs between nodes across time steps, enabling them to capture temporal patterns and maintain internal states. These networks consist of three types of nodes—input nodes that receive external data, output nodes that produce results, and hidden nodes that transform information. RNNs have proven particularly effective for applications requiring temporal understanding. According to the application of this research, RNNs process GOOSE messages as sequences, maintaining a ``memory'' of previous messages~\cite{wang2022anomaly, usmani2022review}. For each message at time $t$, the hidden state update can be given as Eq.~(\ref{hidden_rnn}):

\begin{equation} \label{hidden_rnn}
    h_t = \tanh(W_h \cdot h_{t-1} + W_x \cdot x_t + b)
\end{equation}

where $h_t$ is the current hidden state, $h_{t-1}$ is the previous state, and $x_t$ is the current input. To handle long sequences, a long-short term memory (LSTM) with three gates can be employed as Eq.~(\ref{lstm_rnn}):

\begin{equation} \label{lstm_rnn}
\begin{split}
    \text{Forget gate: } f_t &= \sigma(W_f \cdot [h_{t-1}, x_t] + b_f)\\
    \text{Input gate: } i_t &= \sigma(W_i \cdot [h_{t-1}, x_t] + b_i)\\
    \text{Output gate: } o_t &= \sigma(W_o \cdot [h_{t-1}, x_t] + b_o)
\end{split}
\end{equation}

where $\sigma(z) = 1/(1 + e^{-z})$ is the sigmoid function. Then, the final hidden state is used for classification as the AD process, shown in Eq.~(\ref{anomaly_rnn}):

\begin{equation} \label{anomaly_rnn}
    P(\text{anomaly}) = \sigma(W_{out} \cdot h_T + b_{out})
\end{equation}

\subsubsection{Support Vector Machine (SVM)} 
represents a supervised learning algorithm employed primarily for data classification tasks. The fundamental objective of SVM implementation involves achieving precise categorization of previously unseen data through the optimization of a decision boundary that reduces misclassification rates. This methodology operates through a two-phase process; initially, the algorithm undergoes training using labeled datasets to establish optimal parameters, followed by the application of the trained model to generate class predictions for new, unlabeled instances~\cite{meng2020time, jindal2016decision}. GOOSE messages in IEC61850 networks can be monitored for anomalies using One-Class SVM. Given GOOSE message features $\mathbf{x}$, this algorithm learns a decision boundary around normal GOOSE behavior such that the optimization problem is as Eq.~(\ref{svm_base}):

\begin{equation} \label{svm_base}
\min_{\mathbf{w},\xi,\rho} \frac{1}{2}\|\mathbf{w}\|^2 + \frac{1}{\nu n}\sum_{i=1}^{n}\xi_i - \rho
\end{equation}

subject to: 

\begin{equation} \label{svm_subject}
\begin{split}
\mathbf{w}^T\phi(\mathbf{x}_i) &\geq \rho - \xi_i \\
\xi_i &\geq 0
\end{split}
\end{equation}

The weight vector $\mathbf{w} \in \mathbb{R}^d$ defines the orientation of the separating hyperplane in the feature space, while the offset parameter $\rho$ determines the hyperplane's distance from the origin. The slack variables $\xi_i \geq 0$ enable soft-margin classification by allowing some training points to lie within the margin or on the wrong side of the decision boundary, providing robustness against outliers. The parameter $\nu \in (0,1]$ serves as a user-defined regularization constant that controls the trade-off between maximizing the margin and minimizing the fraction of outliers, effectively determining the upper bound on the fraction of training errors and the lower bound on the fraction of support vectors. The kernel function $K(\mathbf{x}_i, \mathbf{x}_j)$ computes the similarity between data points in the transformed feature space, with the RBF kernel parameter $\gamma > 0$ controlling the influence radius of each support vector; smaller values create smoother decision boundaries and larger values allow more complex, localized boundaries. Finally, the Lagrange multipliers $\alpha_i \geq 0$ determine the contribution of each training sample to the final decision function, with non-zero values identifying the support vectors that define the decision boundary. Using an RBF kernel $K(\mathbf{x}_i, \mathbf{x}_j) = \exp(-\gamma\|\mathbf{x}_i - \mathbf{x}_j\|^2)$, the decision function becomes as Eq.~(\ref{svm_decision_function}). For a new GOOSE message, if $f(\mathbf{x}) < 0$, it is classified as an anomaly.

\begin{equation} \label{svm_decision_function}
f(\mathbf{x}) = \text{sgn}\left(\sum_{i=1}^{n}\alpha_i K(\mathbf{x}_i, \mathbf{x}) - \rho\right)
\end{equation}

\subsubsection{GenAI-based ADSs} leverage self-attention mechanisms to capture complex temporal dependencies in GOOSE message sequences. Given a sequence of GOOSE messages $\mathbf{X} = \{\mathbf{x}_1, \ldots, \mathbf{x}_n\}$, the model learns normal communication patterns through attention-based reconstruction~\cite{vaswani2017attention, gill2023chatgpt}. The architecture consists of an encoder-decoder transformer with positional encoding. The multi-head self-attention mechanism computes as Eq.~(\ref{genai_self_attention}):

\begin{equation} \label{genai_self_attention}
\text{Attention}(Q,K,V) = \text{softmax}\left(\frac{QK^T}{\sqrt{d_k}}\right)V
\end{equation}

where queries $Q$, keys $K$, and values $V$ are linear projections of the GOOSE features. For the AD process, the reconstruction loss can be expressed as Eq.~(\ref{genai_loss}):

\begin{equation} \label{genai_loss}
\mathcal{L}_{rec} = \frac{1}{N}\sum_{i=1}^{N}\|\mathbf{x}_i - \hat{\mathbf{x}}_i\|^2 + \lambda\sum_{h=1}^{H}\text{Entropy}(A_h)
\end{equation}

where $\hat{\mathbf{x}}_i$ is the reconstructed GOOSE message and $A_h$ represents attention weights from head, $h$. The entropy term encourages focused attention patterns. The anomaly score combines reconstruction error and attention anomaly that is represented in Eq.~(\ref{genai_anomaly_score}):

\begin{equation} \label{genai_anomaly_score}
A(\mathbf{x}) = \alpha\|\mathbf{x} - \hat{\mathbf{x}}\|^2 + (1-\alpha)\max_h\text{KL}(A_h\|A_h^{normal})
\end{equation}

This approach excels at detecting temporal anomalies (i.e., unusual message sequences), contextual anomalies (messages inconsistent with historical patterns), and attention-based anomalies where the model's attention deviates from learned normal patterns, providing interpretable detection through attention visualization. The next section shows the results and discussion according to the proposed framework.
\section{Results and Discussion} \label{section-results}
This section presents the results and discussion considering the different layers of the proposed framework, along with the performance evaluation of the ADSs based on a comparison of the proposed methodology with other ML-based methods. Further, three advanced performance evaluation metrics (i.e., informedness, Markedness, and Matthews correlation coefficient - MCC) are provided for a comparison of ADSs in power system applications in addition to standard metrics.
\subsection{Validation of Proposed AATM Methodology for GOOSE Data Generation}
\begin{figure}[!h]
\centerline{\includegraphics[width=1.0\columnwidth]{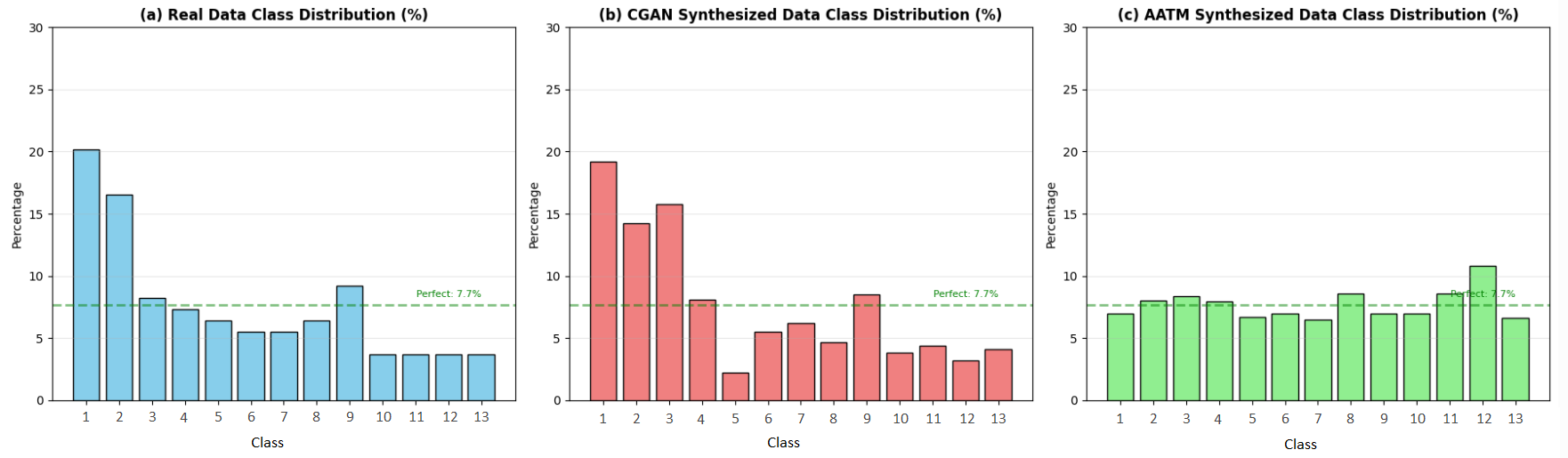}}
\caption{A representation of classes' distributions of GOOSE datasets considering different classes, (a) real data, (b) CGAN synthesized data, and (c) AATM synthesized data.}
\label{CGAN_AATM_synthesized_class}
\end{figure}

The empirical assessment of the proposed AATM methodology demonstrates significant advancements in the capabilities for generating synthetic data when contrasted with the CGAN method. Figure~\ref{CGAN_AATM_synthesized_class} provides an extensive visualization of class distributions spanning three scenarios: the original GOOSE dataset, samples synthesized via the CGAN method, and data generated by the AATM technique.
As can be seen, the original dataset exhibits significant class imbalance, where prevalent categories such as ``Normal'' traffic (approximately 20\%) and data injection (``DI'') attacks ($17\%$) significantly outweigh minority classes, notably ``SP-dataset'' errors which constitute merely 4\% of the total samples. Such distributional disparities present fundamental challenges for developing robust ADSs capable of recognizing diverse attack/error patterns with equal effectiveness. The horizontal axis of the figure shows different classes including the attacks/errors which these classes are enumerated in Table~\ref{tab:distribution_class_AATM_CGAN}.
\begin{table}[!h]
\centering
\caption{A distribution of different classes for CGAN- and AATM-generated GOOSE datasets.}
\label{tab:distribution_class_AATM_CGAN}
\begin{tabular}{|l|c|c|c|c|c|}
\hline
\cellcolor{black} & \textbf{Class} & \textbf{CGAN Count} & \textbf{CGAN \%} & \textbf{AATM Count} & \textbf{AATM \%} \\
\hline
1 & Normal & 961 & 19.2\% & \textbf{350} & \textbf{7.0\%} \\
2 & DI & 712 & 14.2\% & \textbf{401} & \textbf{8.0\%} \\
3 & DOS & 789 & 15.8\% & \textbf{419} & \textbf{8.4\%} \\
4 & RE & 403 & 8.1\% & \textbf{396} & \textbf{7.9\%} \\
5 & SP-time & 111 & 2.2\% & \textbf{334} & \textbf{6.7\%} \\
6 & SP-DM & 277 & 5.5\% & \textbf{348} & \textbf{7.0\%} \\
7 & SP-SM & 311 & 6.2\% & \textbf{325} & \textbf{6.5\%} \\
8 & SP-type & 233 & 4.7\% & \textbf{430} & \textbf{8.6\%} \\
9 & SP-appid & 424 & 8.5\% & \textbf{349} & \textbf{7.0\%} \\
10 & SP-dataset & 193 & 3.9\% & \textbf{348} & \textbf{7.0\%} \\
11 & SP-goid & 219 & 4.4\% & \textbf{428} & \textbf{8.6\%} \\
12 & Packet Loss & 206 & 4.1\% & \textbf{331} & \textbf{6.6\%} \\
13 & Zero-day & 161 & 3.2\% & \textbf{541} & \textbf{10.8\%} \\
\hline
\cellcolor{black} & Total & 5000 & 100.0\% & 5000 & 100.0\% \\
\hline
\end{tabular}
\end{table}
An examination of the synthetic data produced by the CGAN indicates an unforeseen intensification of the existing class imbalance within the original dataset. As demonstrated in Table~\ref{tab:distribution_class_AATM_CGAN}, the CGAN model generates 19.2\% of ``Normal'' class instances, whereas it only produces 2.2\% of ``SP-time'' samples, thereby intensifying the distributional imbalance instead of alleviating it. This occurrence can be attributed to the inherent bias in CGAN models, where the generator network tends to replicate patterns that are dominant in majority classes due to their statistical prominence during the training phase. The ``DOS'' and ``DI'' attack classes also exhibit disproportionate representation at 15.8\% and 14.2\%, respectively, while essential minority classes such as ``SP-dataset'' and ``Zero-day'' attacks remain markedly underrepresented at 3.9\% and 3.2\%, respectively.

\begin{figure}[!h]
\centerline{\includegraphics[width=0.6\columnwidth]{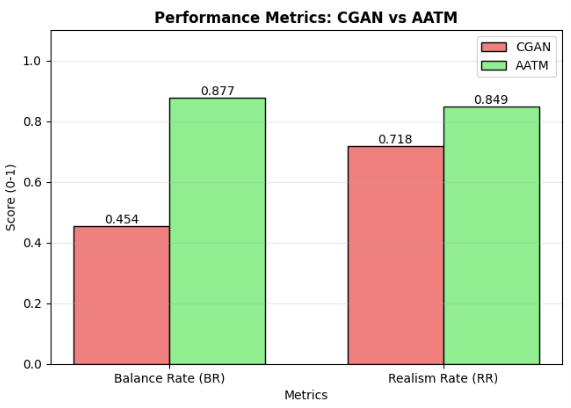}}
\caption{The BR and RR of CGAN- and proposed AATM-generated datasets.}
\label{balance_realism_RR}
\end{figure}

On the other hand, the AATM technique demonstrates exceptional capacity for generating synthetically balanced data distributions. The proposed method achieves remarkable uniformity across all $13$ classes, with representation percentages restricted within a range of 6.5\% to 10.8\%. This balanced generation paradigm is particularly evident in the transformation of traditionally infrequent categories: ``SP-time'' classes increase from 2.2\% under CGAN to 6.7\% with the AATM method, while ``Zero-day'' attacks experience a substantial enhancement from 3.2\% to 10.8\%. Simultaneously, overly dominant classes undergo appropriate reduction, with ``Normal'' traffic decreasing from 19.2\% to 7.0\%, thereby contributing to overall distributional balance. These distribution percentages show that the CGAN technique generated more ``Normal'' classes as it could not get all the correct patterns for this class. Also, some of the attacks/errors were mistakenly generated as they could be the ``Zero-day'' attacks. The quantitative assessment presented in Fig.~\ref{balance_realism_RR} further validates the superiority of the AATM methodology through two critical metrics. 
The BR, an index that quantifies the distributional uniformity across generated samples, demonstrates a notable increase from 0.454 in CGAN to 0.877 in AATM, reflecting a 93\% enhancement in class balance. This substantial improvement underscores the AATM's effectiveness in addressing the core challenge of generating representative samples across all classes, including attacks/errors and normal data, regardless of their frequency in the training dataset. Furthermore, RR, which evaluates the credibility and quality of synthesized samples, shows an advancement from 0.718 to 0.849, signifying an 18\% enhancement in the fidelity of generated data. This simultaneous progress in both BR and RR underscores the AATM's capability in producing high-quality synthetic samples while preserving class balance.
The implications of these findings reach far beyond simple enhancements in statistical metrics, providing substantial advantages for real-world cybersecurity implementations. ADSs that are trained on datasets with unbalanced distributions often demonstrate a reduced ability to accurately identify rare yet possibly disastrous attack paths. By generating balanced synthetic datasets, the AATM technique facilitates the creation of detection models that maintain an equitable performance across a wide array of threat signatures, achieving similar levels of detection accuracy. This characteristic proves particularly valuable for zero-day attack detection, where AATM's generation of 10.8\% samples compared to CGAN's 3.2\% provides sufficient training instances for models to develop robust recognition capabilities for these critical yet infrequent threats. Despite these findings, it is important to evaluate certain limitations in the context of interpreting the results. The observed residual variation in class percentages produced by the AATM algorithm, which varies between 6.5\% and 10.8\%, illustrates that achieving an entirely uniform distribution remains an ongoing challenge yet to be completely resolved. Future research could investigate the scalability attributes of the AATM technique when applied to datasets with an expanded number of classes. Additionally, these studies might assess the feasibility of integrating constraints specific to particular domains to not only enhance the realism of samples but also to ensure a balanced distribution across the dataset.
\subsection{Performance Evaluation Metrics in an AD Process}
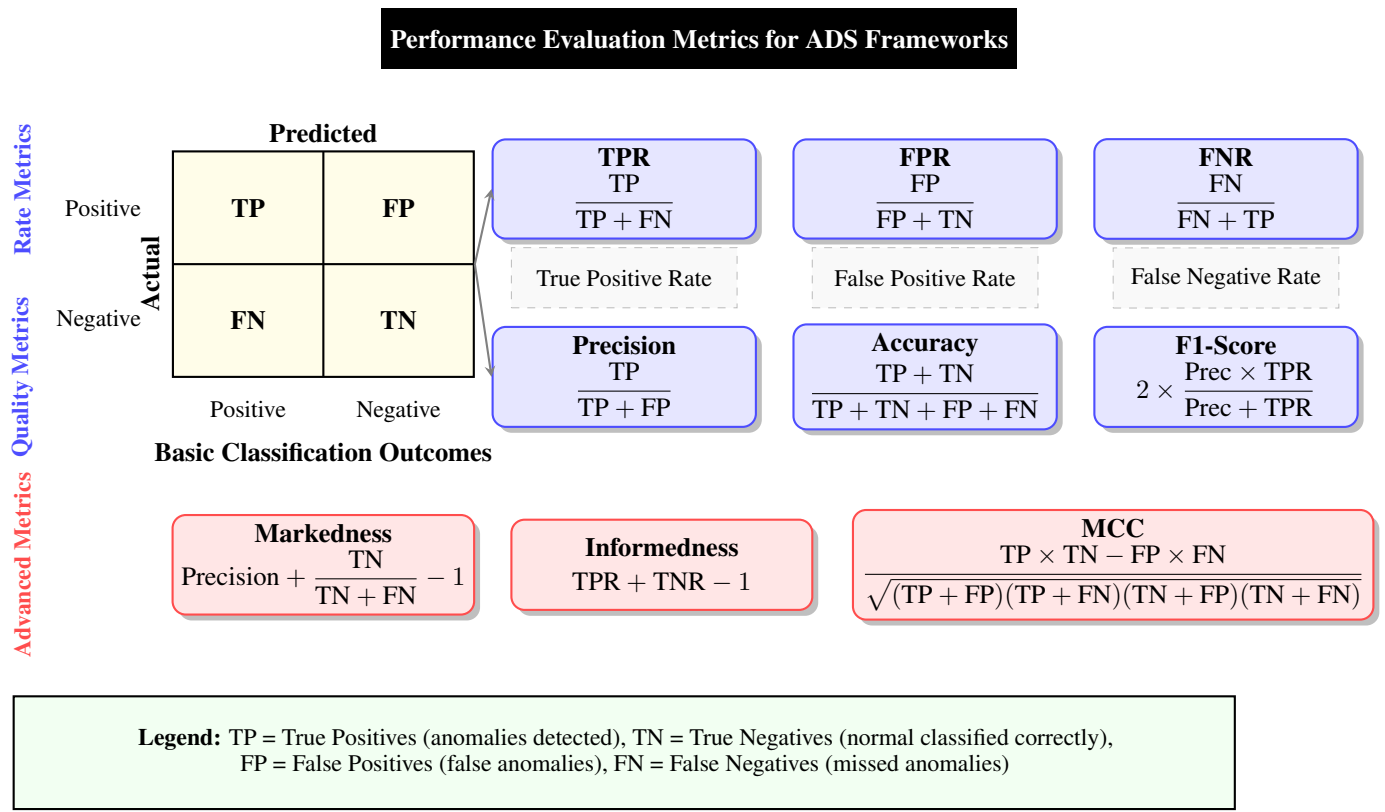
\begin{figure*}[ht]
\centering
\begin{tikzpicture}[
    box/.style={rectangle, draw=black, thick, minimum width=2.5cm, minimum height=1cm, align=center, fill=white, drop shadow},
    metricbox/.style={rectangle, draw=blue!70, thick, minimum width=3.5cm, minimum height=1.2cm, align=center, fill=blue!10, rounded corners=5pt, drop shadow},
    advbox/.style={rectangle, draw=red!70, thick, minimum width=4cm, minimum height=1.2cm, align=center, fill=red!10, rounded corners=5pt, drop shadow},
    formula/.style={rectangle, draw=gray!50, dashed, minimum width=3cm, minimum height=0.8cm, align=center, fill=gray!5},
    arrow/.style={->, >=stealth, thick, gray},
    title/.style={rectangle, fill=black, text=white, font=\bfseries, minimum width=4cm, minimum height=0.8cm, align=center}
]

\node[title] at (-1, 8) {Performance Evaluation Metrics for ADS Frameworks};

\node[draw, thick, minimum width=4cm, minimum height=3cm, fill=yellow!10] (matrix) at (-6, 5) {};
\draw[thick] (-8, 5) -- (-4, 5);
\draw[thick] (-6, 3.5) -- (-6, 6.5);

\node at (-7, 5.75) {\textbf{TP}};
\node at (-5, 5.75) {\textbf{FP}};
\node at (-7, 4.25) {\textbf{FN}};
\node at (-5, 4.25) {\textbf{TN}};

\node[above] at (-6, 6.5) {\textbf{Predicted}};
\node[rotate=90, left] at (-8.3, 5.5) {\textbf{Actual}};

\node[below] at (-7, 3.3) {\small Positive};
\node[below] at (-5, 3.3) {\small Negative};
\node[left] at (-8.3, 5.75) {\small Positive};
\node[left] at (-8.3, 4.25) {\small Negative};

\node[font=\bfseries] at (-6, 2.5) {Basic Classification Outcomes};

\node[metricbox] (tpr) at (-2, 6) {\textbf{TPR}\\[2pt]$\dfrac{\text{TP}}{\text{TP}+\text{FN}}$};
\node[metricbox] (fpr) at (2, 6) {\textbf{FPR}\\[2pt]$\dfrac{\text{FP}}{\text{FP}+\text{TN}}$};
\node[metricbox] (fnr) at (6, 6) {\textbf{FNR}\\[2pt]$\dfrac{\text{FN}}{\text{FN}+\text{TP}}$};

\node[metricbox] (precision) at (-2, 3.5) {\textbf{Precision}\\[2pt]$\dfrac{\text{TP}}{\text{TP}+\text{FP}}$};
\node[metricbox] (accuracy) at (2, 3.5) {\textbf{Accuracy}\\[2pt]$\dfrac{\text{TP}+\text{TN}}{\text{TP}+\text{TN}+\text{FP}+\text{FN}}$};
\node[metricbox] (f1) at (6, 3.5) {\textbf{F1-Score}\\[2pt]$2 \times \dfrac{\text{Prec} \times \text{TPR}}{\text{Prec} + \text{TPR}}$};

\node[advbox] (markedness) at (-6, 1) {\textbf{Markedness}\\[2pt]$\text{Precision} + \dfrac{\text{TN}}{\text{TN}+\text{FN}}-1$};
\node[advbox] (informedness) at (-1.5, 1) {\textbf{Informedness}\\[2pt]$\text{TPR}+\text{TNR}-1$};
\node[advbox] (mcc) at (4.5, 1) {\textbf{MCC}\\[2pt]$\dfrac{\text{TP} \times \text{TN} - \text{FP} \times \text{FN}}{\sqrt{(\text{TP}+\text{FP})(\text{TP}+\text{FN})(\text{TN}+\text{FP})(\text{TN}+\text{FN})}}$};

\draw[arrow] (matrix.east) -- (tpr.west);
\draw[arrow] (matrix.east) -- (precision.west);

\node[rotate=90, font=\footnotesize\bfseries, text=blue!70] at (-10, 6) {Rate Metrics};
\node[rotate=90, font=\footnotesize\bfseries, text=blue!70] at (-10, 3.5) {Quality Metrics};
\node[rotate=90, font=\footnotesize\bfseries, text=red!70] at (-10, 1) {Advanced Metrics};

\node[formula, below=0.1cm of tpr] {\footnotesize True Positive Rate};
\node[formula, below=0.1cm of fpr] {\footnotesize False Positive Rate};
\node[formula, below=0.1cm of fnr] {\footnotesize False Negative Rate};

\node[draw, thick, minimum width=10cm, minimum height=1.5cm, fill=green!5] at (-2, -1.5) {
\begin{minipage}{16cm}
\centering
\footnotesize
\textbf{Legend:} TP = True Positives (anomalies detected), TN = True Negatives (normal classified correctly),
\\ FP = False Positives (false anomalies), FN = False Negatives (missed anomalies)
\end{minipage}
};

\end{tikzpicture}
\caption{Performance evaluation metrics and their mathematical formulations for assessing ADS frameworks. The confusion matrix (left) shows basic classification outcomes, which are used to derive rate metrics (blue boxes) and advanced evaluation metrics (red boxes).}
\label{EvalMetric_1}
\end{figure*}

In this section, a comparative analysis is conducted to assess the performance and effectiveness of the proposed GenAI-based ToD framework over ML-based ADSs. Fig.~\ref{EvalMetric_1} shows the different descriptions and formulations for standard and advanced evaluation metrics to make a comparison between these frameworks. 
These metrics represent essential evaluation criteria for ADSs implementing security monitoring of GOOSE messages. The fundamental classification metrics include the true positives (TP), true negatives (TN), false positives (FP), and false negatives (FN) indicators. In this context, these primary metrics quantify correctly identified anomalies, properly classified normal traffic, incorrectly flagged normal communications, and undetected anomalous events, respectively. The true positive rate (TPR), calculated as $\frac{TP}{TP+FN}$, quantifies the system's sensitivity in detecting actual anomalies. Conversely, the false positive rate (FPR), expressed as $\frac{FP}{FP+TN}$, measures the likelihood of false alarms, while the false negative rate (FNR), determined by $\frac{FN}{TP+FN}$, evaluates detection failures, a particularly critical metric for security applications in industrial control systems. The Precision, formulated as $\frac{TP}{TP+FP}$, indicates the reliability of positive predictions, while accuracy, calculated as $\frac{TP+TN}{TP+TN+FP+FN}$, assesses overall classification correctness across the entire messages collection. The F1-Score, expressed as $2 \times \frac{Precision \times TPR}{Precision+TPR}$, provides a harmonic mean of precision and TPR, offering a balanced assessment when both false alarms and missed detections possess significant operational implications in substations.

Within this work, advanced metrics such as Markedness, Informedness, and the MCC are utilized to evaluate the consistency, decision-making precision, and classification quality, ranging from $-1$ to $1$. These metrics prove beneficial in the AD process to assess the efficiency and applicability of frameworks. Specifically, in the context of AD applied to GOOSE datasets, Markedness is indispensable for measuring the model's capability to reduce both false positives (FPs) and false negatives (FNs). A high Markedness value signifies a robust AD framework that minimizes erroneous alerts, thereby ensuring stability and optimal performance at substations by decreasing unnecessary disruptions. Meanwhile, Informedness measures the model's proficiency in recognizing dataset pattern variations that indicate anomalies. Particularly for AD scenarios using GenAI models — where real anomalies may occur infrequently — MCC is advantageous as it ensures the model's performance reflects its true efficacy and is not excessively affected by a larger class size~\cite{de2022general}.

\subsection{A Comparative Analysis of the GenAI-based and ML-based ADSs}
\begin{figure}[!h]
    \centering
    \begin{subfigure}[b]{0.45\textwidth}
        \includegraphics[width=\textwidth]{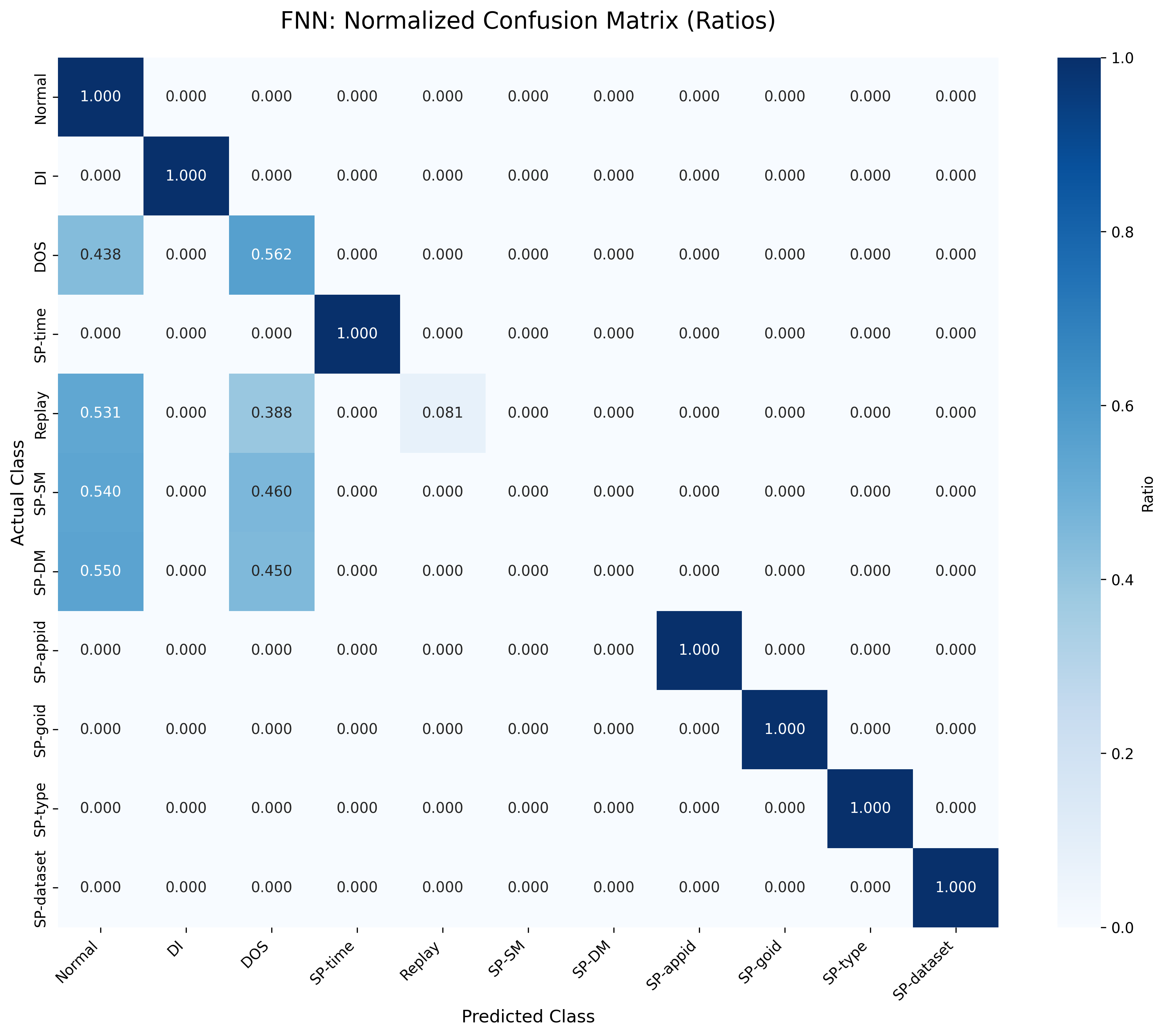}
        \caption{}
        \label{subfig_confusion_FNN_normalized}
    \end{subfigure}
    \hfill
    \begin{subfigure}[b]{0.45\textwidth}
        \includegraphics[width=\textwidth]{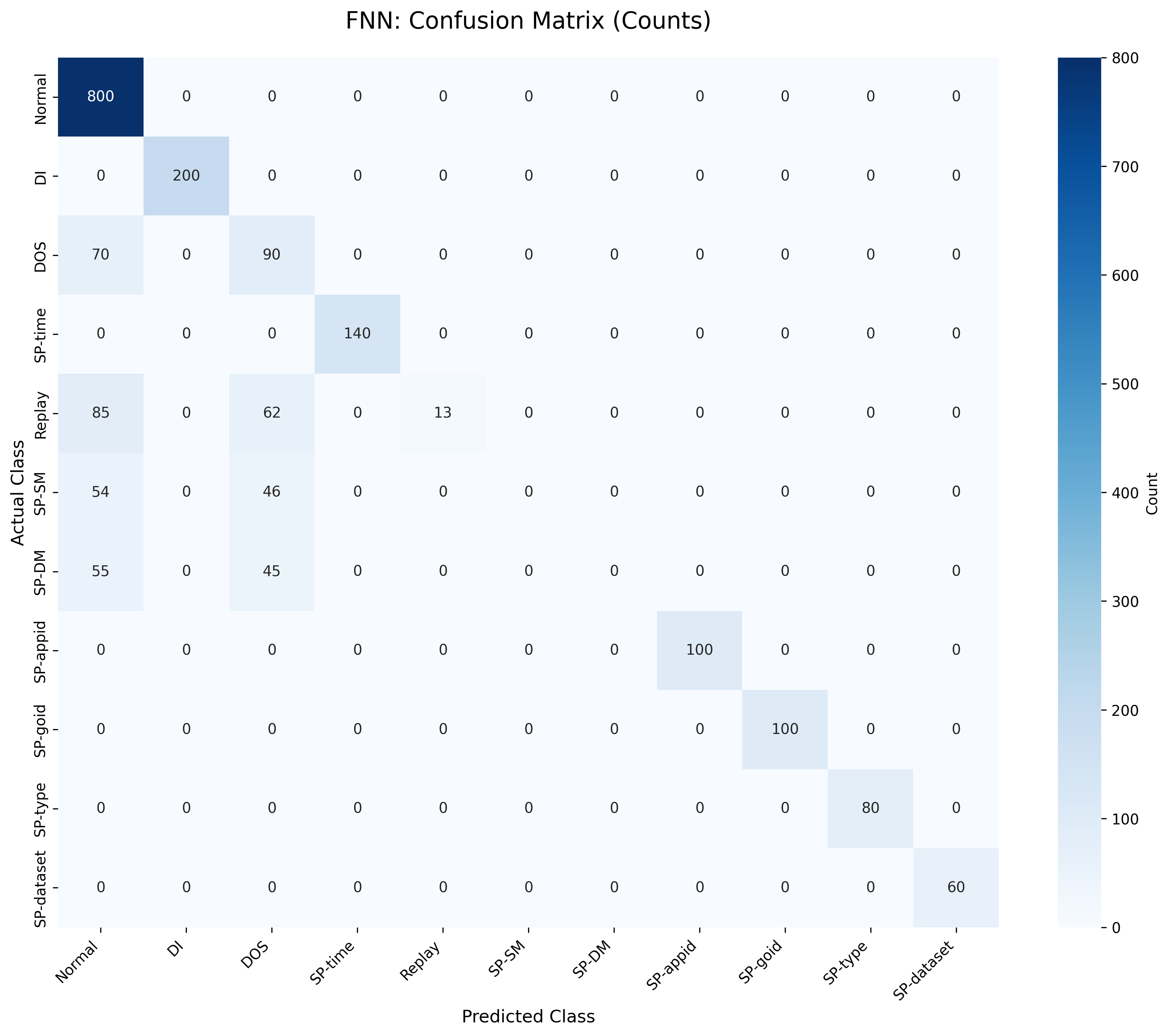}
        \caption{}
        \label{subfig_confusion_FNN_counts}
    \end{subfigure}
    \caption{Confusion matrices of an FNN-based ADS, trained by the proposed AATM-generated GOOSE datasets, (a) normalized ratios (b) counts.}
    \label{fig_confusion_FNN}
\end{figure}

\begin{figure}[!h]
    \centering
    \begin{subfigure}[b]{0.45\textwidth}
        \includegraphics[width=\textwidth]{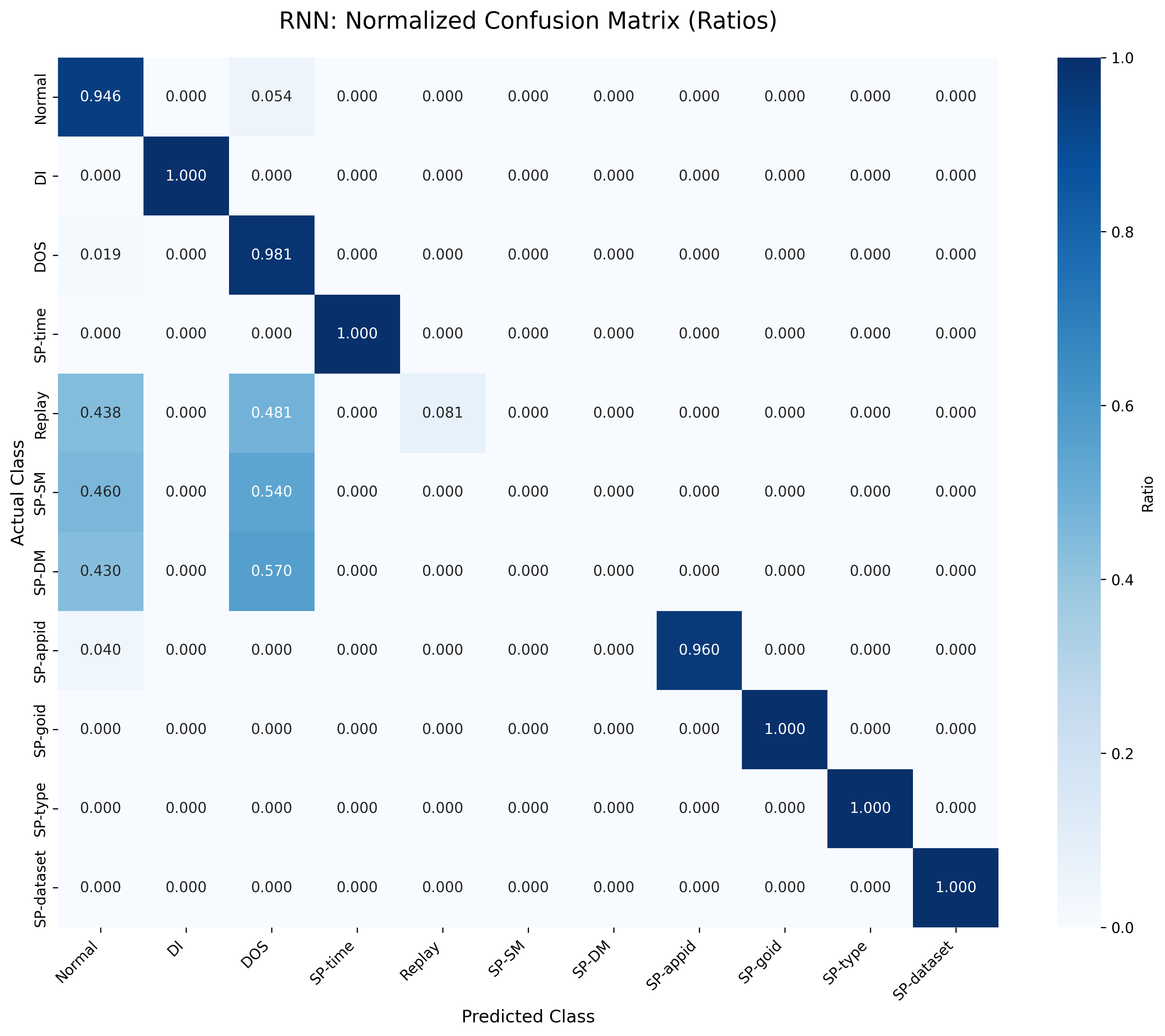}
        \caption{}
        \label{subfig_confusion_RNN_normalized}
    \end{subfigure}
    \hfill
    \begin{subfigure}[b]{0.45\textwidth}
        \includegraphics[width=\textwidth]{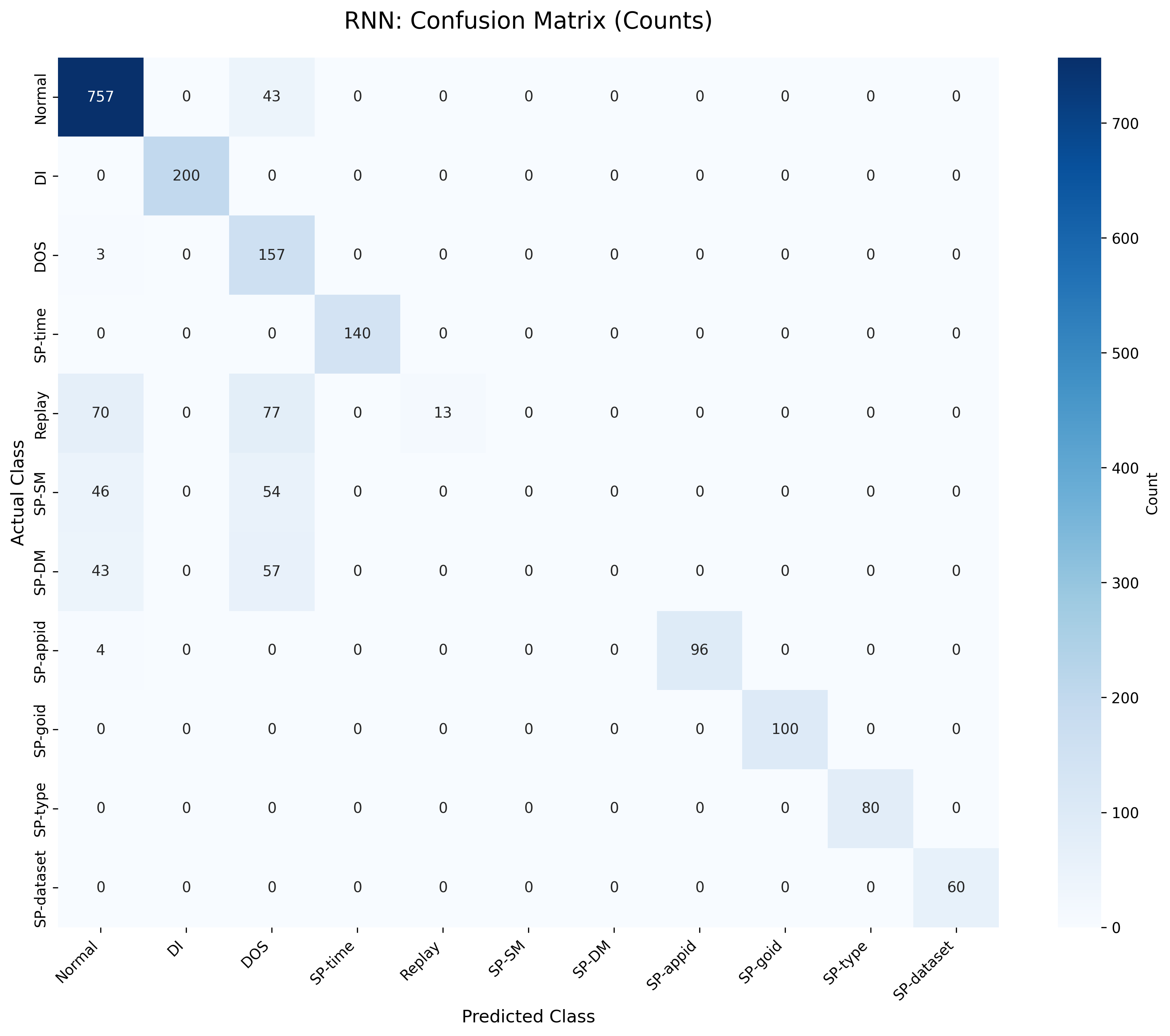}
        \caption{}
        \label{subfig_confusion_RNN_counts}
    \end{subfigure}
    \caption{Confusion matrices of an RNN-based ADS, trained by the proposed AATM-generated GOOSE datasets, (a) normalized ratios (b) counts.}
    \label{fig_confusion_RNN}
\end{figure}

\begin{figure}[!h]
    \centering
    \begin{subfigure}[b]{0.45\textwidth}
        \includegraphics[width=\textwidth]{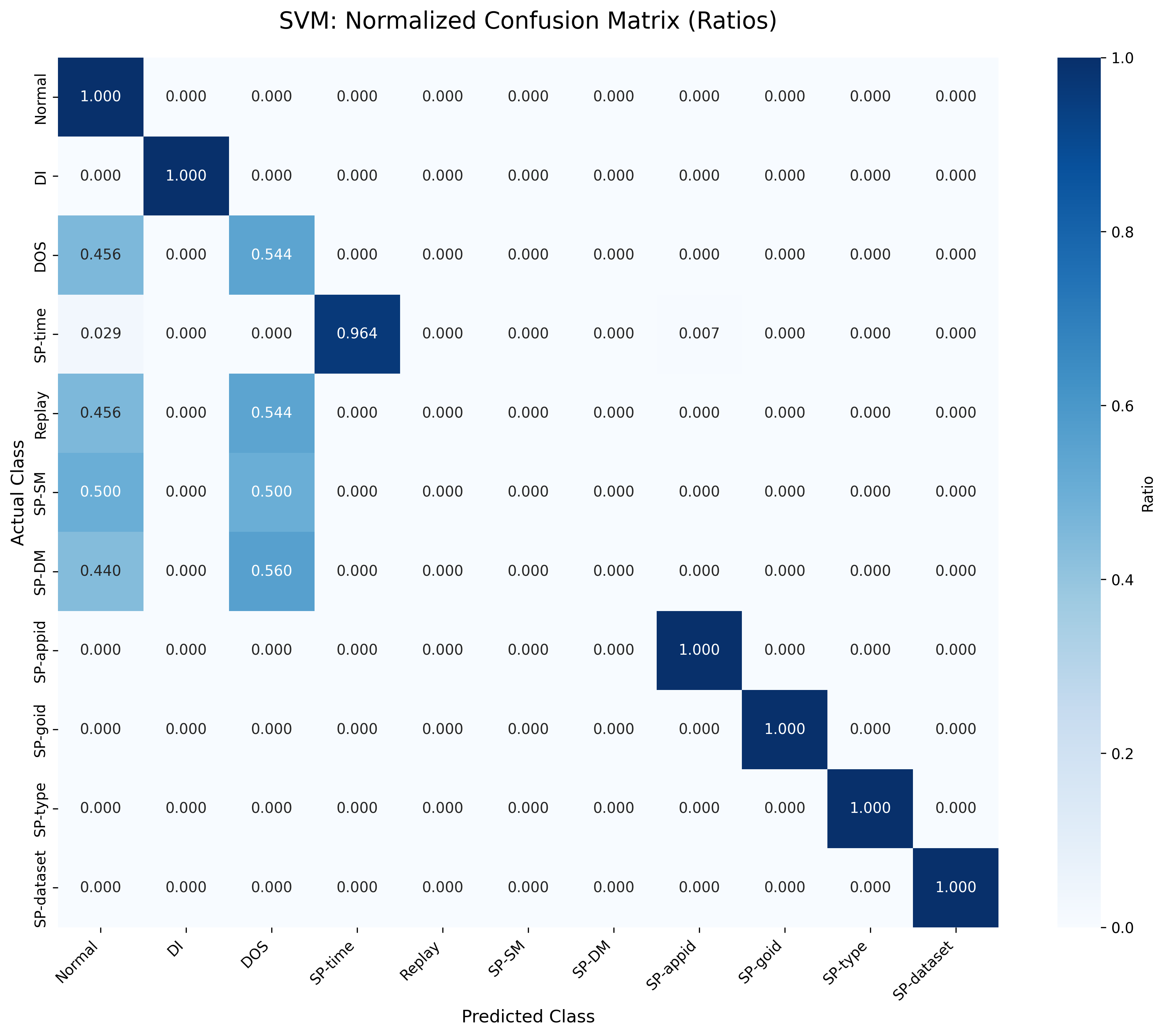}
        \caption{}
        \label{subfig_confusion_SVM_normalized}
    \end{subfigure}
    \hfill
    \begin{subfigure}[b]{0.45\textwidth}
        \includegraphics[width=\textwidth]{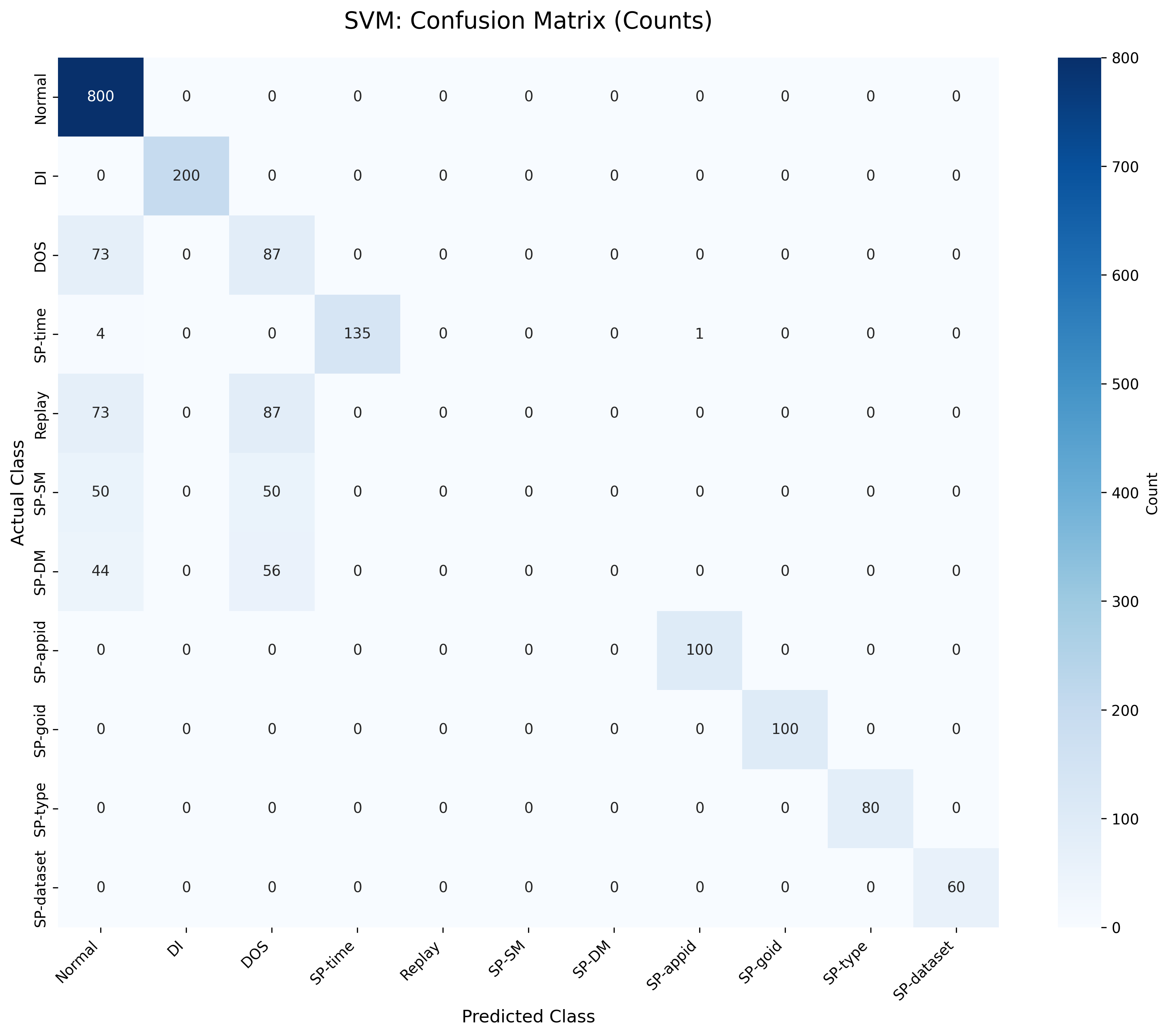}
        \caption{}
        \label{subfig_confusion_SVM_counts}
    \end{subfigure}
    \caption{Confusion matrices of an SVM-based ADS, trained by the proposed AATM-generated GOOSE datasets, (a) normalized ratios (b) counts.}
    \label{fig_confusion_SVM}
\end{figure}

\begin{figure}[!h]
    \centering
    \includegraphics[width=0.5\textwidth]{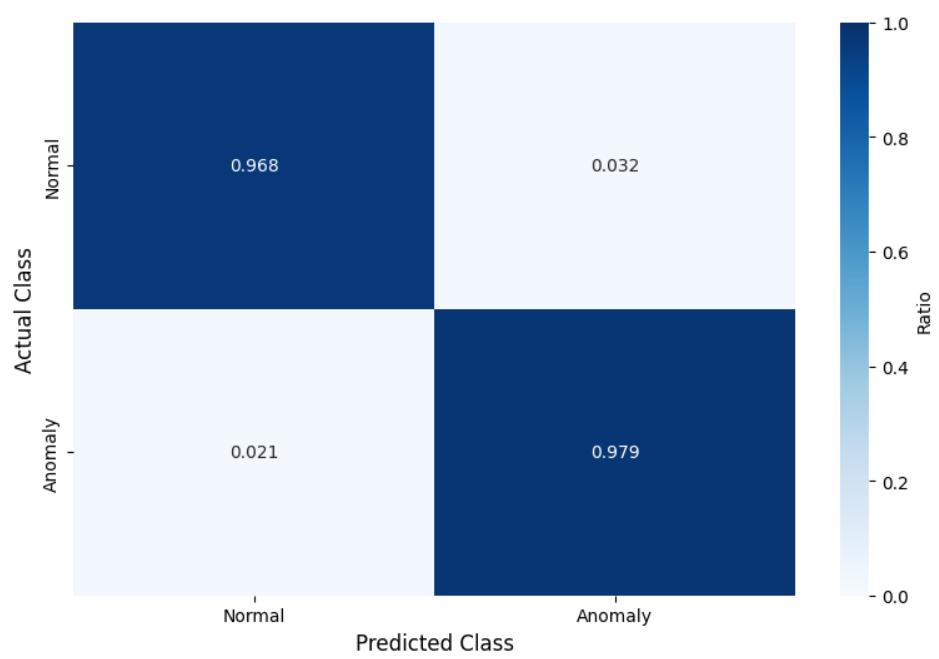}
    \caption{A normalized confusion matrix of the GenAI-based ADS, trained by the proposed AATM-generated GOOSE datasets considering the normal and anomalous classes.}
    \label{GenAI_CM_normalized_2x2}
\end{figure}

\begin{figure}[!h]
    \centering
    \includegraphics[width=0.8\textwidth]{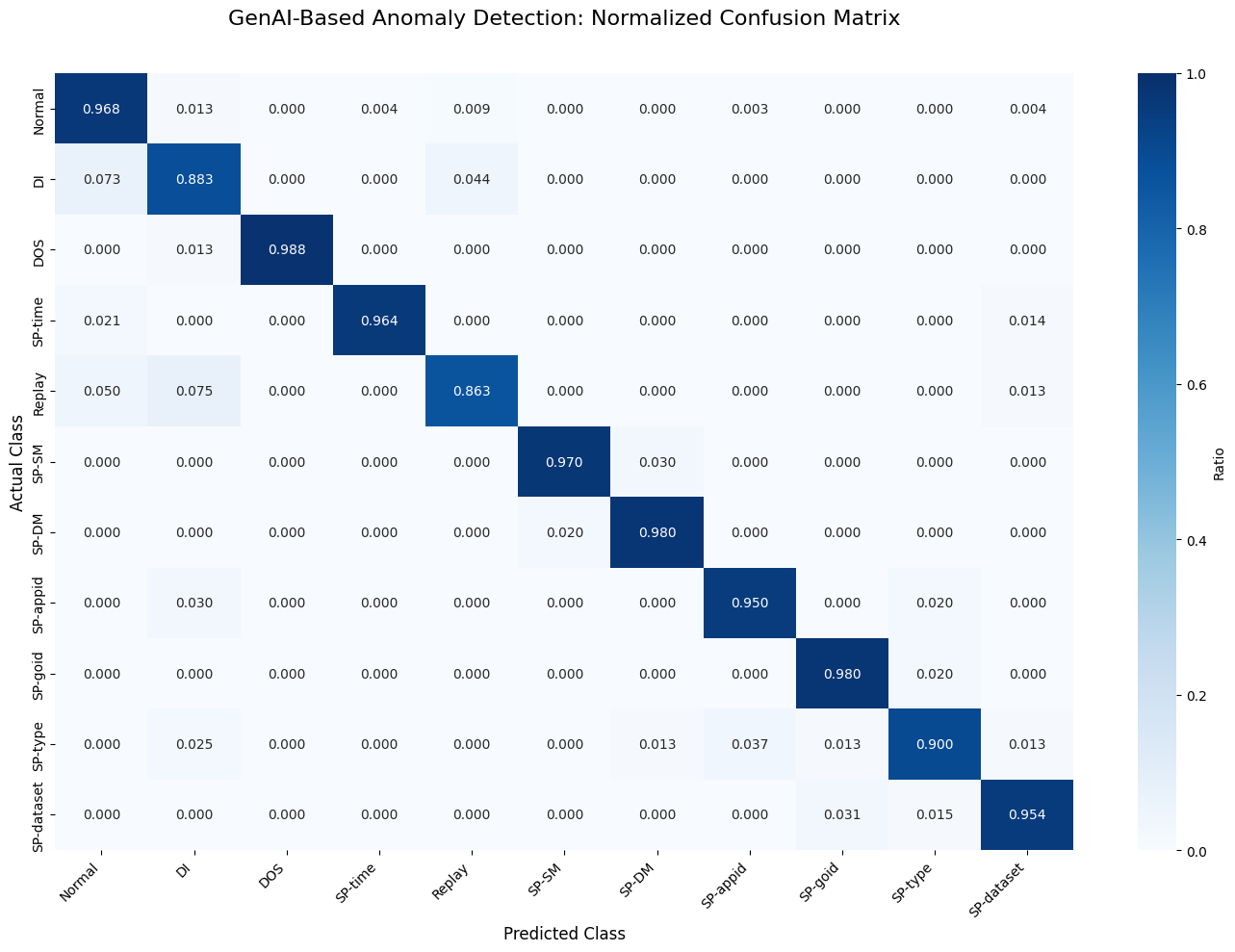}
    \caption{A normalized confusion matrix of the GenAI-based ADS, trained by the proposed AATM-generated GOOSE datasets for all classes.}
    \label{GenAI_CM_normalized}
\end{figure}

\begin{figure}[!h]
    \centering
    \includegraphics[width=0.8\textwidth]{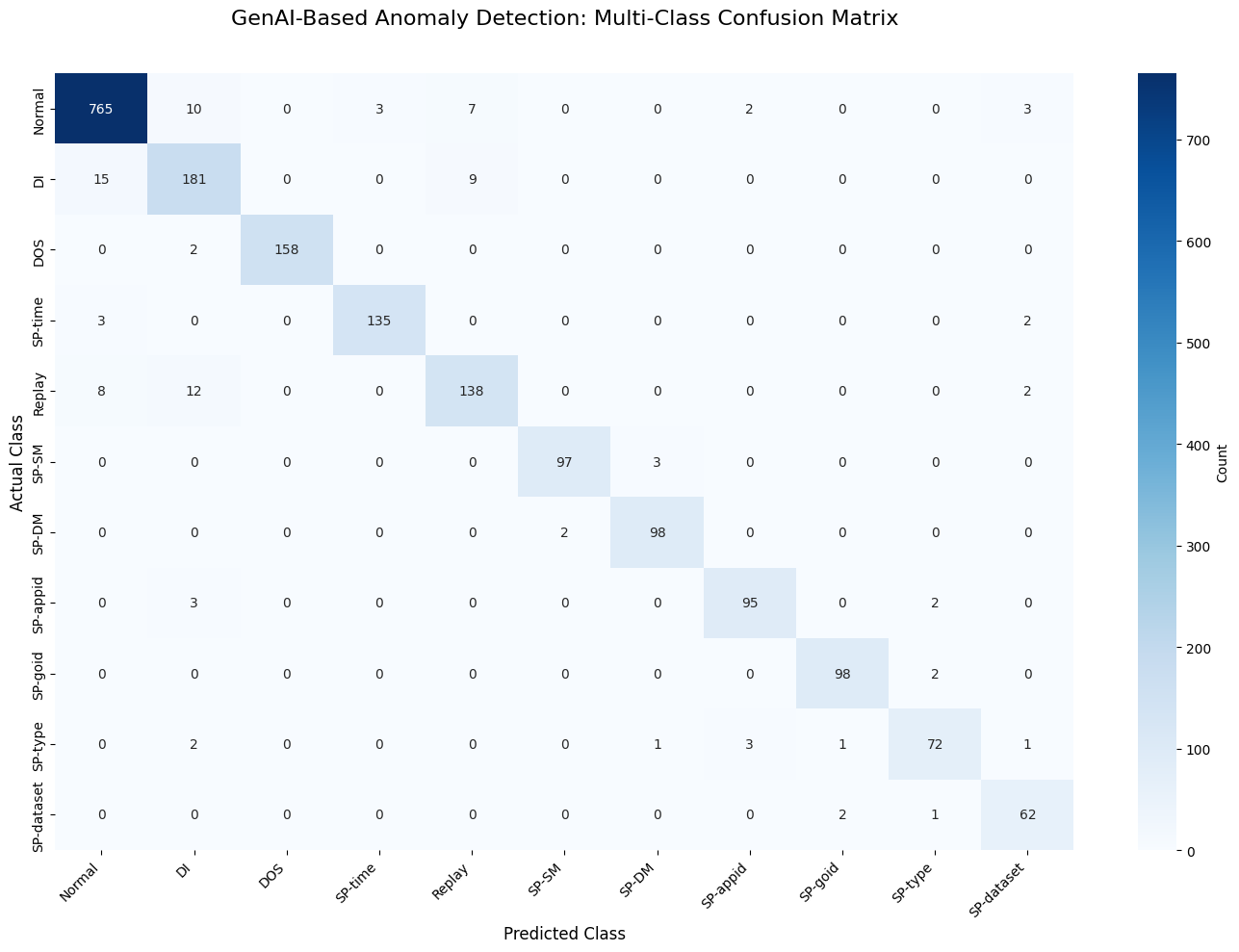}
    \caption{A confusion matrix (counts) of the GenAI-based ADS, trained by the proposed AATM-generated GOOSE datasets for all classes.}
    \label{GenAI_CM_counts}
\end{figure}

This section presents a comparative analysis of four ADSs considering GOOSE datasets generated using the proposed AATM technique. The comparison includes three ML algorithms — FNN, RNN, SVM, in addition to a GenAI-based ADS. The empirical results provide robust evidence of the superior effectiveness exhibited by the GenAI-based ADS, especially when evaluated against ML models. It is presented in Table~\ref{table:GenAI_ML_comparison} that the proposed GenAI approach achieves an outstanding classification accuracy rate of 97.9\%. 
\begin{table}[!h]
\centering
\caption{A comparison of GenAI- and ML-based ADSs using AATM-generated GOOSE datasets.}
\label{table:GenAI_ML_comparison}
\begin{tabular}{|c|c|c|c|c|}
\hline
{\textbf{Algorithms}} & {FNN} & {RNN} & {SVM} & {Anthropic Claude Pro (GenAI-based ADS)} \\
\hline
\makecell{\textbf{Standard Metrics}} & \cellcolor{black} & \cellcolor{black} & \cellcolor{black} & \cellcolor{black} \\
\hline
\makecell{\textit{TPR}} & 79\% & 87.9\% & 79.1\%  & \textbf{97.9\%}  \\
\hline
\makecell{\textit{FPR}} & \textbf{0\%} & 10.6\% & \textbf{0\%} & 3.2\% \\
\hline
\makecell{\textit{FNR}} & 21\% & 12.08\% & 20.9\%  & \textbf{2.1\%} \\
\hline
\makecell{\textit{Precision}} & \textbf{100\%} & 92.5\% & \textbf{100\%} & 97.9\% \\
\hline
\makecell{\textit{Accuracy}} & 87.4\% & 88.5\% & 87.4\% & \textbf{97.5\%} \\
\hline
\makecell{\textit{F1-Score}} & 88.3\% & 90.2\% & 88.3\% & \textbf{97.9\%} \\
\hline
\makecell{\textbf{Advanced Metrics}} & \cellcolor{black} & \cellcolor{black} & \cellcolor{black} & \cellcolor{black} \\
\hline
\makecell{\textit{Markedness}} & 0.76 & 0.756 & 0.761 & \textbf{0.947} \\
\hline
\makecell{\textit{Informedness}} & 0.79 & 0.773 & 0.791 & \textbf{0.947} \\
\hline
\makecell{\textit{MCC}} & 0.775 & 0.764 & 0.776 & \textbf{0.945} \\
\hline
\end{tabular}
\end{table}
This table demonstrates a significant enhancement in performance metrics compared to ML models, such as FNN, which achieves an accuracy rate of 87.4\%, RNN achieving 88.5\%, and SVM, also at 87.4\%. The observed improvement (approximately a 10 percentage point increase in the accuracy metric) represents a significant advancement in the domain of GOOSE messages monitoring. Further, this proposed framework showed better performance in almost all other metrics. A thorough analysis of the confusion matrices demonstrated in Figs.~\ref{fig_confusion_FNN} --~\ref{GenAI_CM_counts} offers a comprehensive understanding of the classification features displayed by different approaches. The GenAI framework's confusion matrices, i.e.,  Figs.~\ref{GenAI_CM_normalized_2x2}~--~\ref{GenAI_CM_counts} demonstrate prominent diagonal concentration with negligible inter-class confusion, reflecting robust differentiating capabilities. Notably remarkable is the normalized confusion matrix of the GenAI implementation (Fig.~\ref{GenAI_CM_normalized}), which showcases near-unity values along the diagonal for the majority of classes, contrasting markedly with the substantial class overlap observed in ML approaches, especially when distinguishing between structurally similar class signatures.

The performance metric analysis indicates that although the SVM algorithm achieves a comparable FPR of 3.2\% to the GenAI model, the latter distinguishes itself by simultaneously maintaining the minimal FNR of 2.1\%, substantially lower than the 21\%, 12.08\%, and 20.9\% observed for FNN, RNN, and SVM approaches, respectively. Such balanced error characteristics prove essential for practical deployment scenarios where both false alarms and undetected threats must be minimized. Additional performance metrics meticulously confirm the superiority of the GenAI framework. This system achieves Markedness and Informedness values reaching 0.947, a notable advancement well beyond the ML benchmarks typically observed between 0.76 and 0.791. Furthermore, it performs well in accurately classifying classes, effectively distinguishing between situations indicative of attack and those that are non-malicious, thereby demonstrating a high degree of precision in its operation. Furthermore, the MCC, recording at 0.945, provides additional validity to the comprehensive effectiveness of the model. This coefficient reflects an enhancement of approximately 20\% compared to standard baseline methodologies, confirming GenAI’s advanced capabilities and distinguished performance.

Analysis of the multi-class confusion matrix (i.e., Fig.~\ref{GenAI_CM_counts}) clarifies the GenAI framework's classification capabilities across diverse attack/error categories. The system demonstrates remarkable competence in distinguishing subtle attack variations including DI, DOS, temporal anomalies (i.e., SP-time), feature-based manipulations (SP-[feature]), and RE attacks, with negligible inter-category confusion. This refined detection resolution originates from the framework's inherent capacity for semantic interpretation of message dynamics, as illustrated in the diagnostic output where specific message deviations and contextual anomalies are clearly identified. The GenAI model possesses an advanced capacity for semantic understanding, which facilitates its detection of complex attack/error patterns that are often missed by traditional statistical approaches. This performance is quantitatively supported by its Markedness and Informedness scores, both achieving a value of 0.947. The model's advantage in this domain stems from its context-based reasoning abilities concerning message violations, differing markedly from methods that depend solely on numerical pattern recognition. According to this concept, a textual portion of the ``Response'' for GenAI-based ADS in Anthropic Claude Pro is presented in the following box for more clarification.
\begin{responsebox}
\begin{enumerate}
    \item \textbf{Semantic Pattern Understanding Phase}

\noindent I would analyze each dataset by:
\begin{itemize}
    \item \textbf{Reading the temporal flow} of messages within each dataset
    \item \textbf{Understanding the protocol semantics} - recognizing that GOOSE messages follow specific IEC61850 patterns
    \item \textbf{Contextual interpretation} of field relationships (DM/SM pairs, stNum/sqNum)
\end{itemize}

\item \textbf{Rule-Based Reasoning with Contextual Intelligence}

\noindent Unlike rigid rule engines, I would apply the 8 rules with contextual understanding:

\noindent \textbf{Rule Application Strategy:}
\begin{itemize}
    \item \textbf{Rule 1 \& 3:} Track sqNum progression and stNum monotonicity within DM/SM contexts
    \item \textbf{Rule 2 \& 8:} Understand the semantic relationship between data changes and sqNum
    \item \textbf{Rule 4:} Detect configuration changes in categorical fields
    \item \textbf{Rule 5-7:} Analyze temporal patterns for DOS and timing anomalies
\end{itemize}
\item \textbf{Multi-Level Anomaly Classification Process}

\noindent \textbf{Level 1 - Pattern Recognition:}

\noindent \texttt{For each dataset:}

\noindent \texttt{1. Identify message sequence patterns}

\noindent \texttt{2. Detect temporal anomalies (ms-level for DOS, second-level for SP-time)}

\noindent \texttt{3. Recognize protocol violations}

\noindent \textbf{Level 2 - Semantic Classification:}
\begin{itemize}
    \item \textbf{Normal:} All rules satisfied, expected GOOSE behavior
    \item \textbf{DI:} StNum changes without proper data correlation
    \item \textbf{DOS:} $>$ 10 messages within 10 microseconds
    \item \textbf{SP-time:} $> $10 second gaps between messages
    \item \textbf{SP-[feature]:} Unexpected changes in specific fields
    \item \textbf{RE:} Repeated sequence patterns
\end{itemize}

\noindent \textbf{Level 3 - Contextual Validation:}
\begin{itemize}
    \item Cross-reference patterns with known GOOSE protocol behavior
    \item Apply semantic understanding of industrial control system operations
\end{itemize}

\noindent \textbf{Example Analysis Output:}

\noindent \texttt{Dataset \#1: ANOMALY (DI Class)}

\noindent \texttt{Reasoning: Detected stNum change from 27 to 28 at row 5 without corresponding data1/data2 change. This violates the expected GOOSE protocol behavior where stNum changes should correlate with data modifications.}
\vspace{2mm}

\noindent \texttt{Dataset \#2: NORMAL}

\noindent \texttt{Reasoning: All sequences follow expected patterns. sqNum increments properly (150$\rightarrow$151$\rightarrow$152), stNum remains constant at 27, no timing anomalies detected.}
\vspace{2mm}

\noindent \texttt{Dataset \#3: ANOMALY (DOS Class)}

\noindent \texttt{Reasoning: Identified 12 messages within 8 microseconds (rows 3-14), indicating a DOS pattern with abnormally rapid message transmission.}
\end{enumerate}
\end{responsebox}
As can be observed, it can follow the provided rules in addition to the patterns in the datasets to provide a suitable reason according to each class. Some examples of outputs based on three datasets are represented at the end of the box.
\section{Conclusions and Future Work} \label{conclusion-section}
This section generally presents the AATM technique for balanced and realistic data generation, novel GenAI-based ADSs, and an enhanced performance of GenAI-based ADSs over ML-based ADSs. Initially, because there are insufficient datasets of IEC61850-based communications as well as the generation of realistic zero-day attacks, this research proposes a novel pre-processing technique known as AATM for data generation which is perturbation- and mutation-based to enhance the RR and BR of the generated synthesized datasets, which show enhancements over another pre-processing technique. Secondly, a GenAI-based ToD ADS is presented in IEC61850-based communication messages in digital substations which rather outperforms the traditional ML-based ADSs in terms of no necessity for the re-training process, less effort, and the ability of analysis of categorical features in multi-cast messages. Then, considering the generated datasets and suggested GenAI-based ADS, the performance of this ADS is assessed based on standard and advanced performance metrics. According to the results, it demonstrates that the GenAI-based ToD framework implemented by Anthropic Claude Pro has superior performance compared with other ML models. 
Finally, the system's capacity for semantic comprehension, demonstrated through its proficiency in contextualizing message/pattern anomalies and detecting sophisticated attack/error signatures undetected by solely quantitative approaches, represents a qualitative evolution beyond traditional pattern recognition models. The framework's consistent high performance across diverse normal/threat vectors, including data manipulation, operational issues, temporal attacks, and message RE scenarios, in combination with its inherent scalability and sustainability features, validates its suitability for production deployment. These findings position GenAI as a crucial technology in securing critical infrastructure, providing a resilient, explainable, and evolutionary security solution capable of addressing emergent threats within digital substation environments while preserving the rigorous reliability requirements fundamental to electrical grid operations. 

The development of security frameworks for substations opens up numerous promising future pathways. Present detection techniques must evolve to include the entire suite of IEC61850 communication protocols. This extension effort results in a comprehensive monitoring ecosystem, surpassing the constraints associated with single-protocol analysis. Tailored intelligent systems can be deployed within utility-managed infrastructure using community-developed language models. These deployments establish self-contained operational environments that satisfy rigorous regulatory compliance requirements. Power system physics-aware computational models facilitate instantaneous threat identification within these frameworks. This architectural approach naturally extends toward interconnected substation networks. Mechanisms for cryptographically protected information exchange facilitate joint threat evaluation among installations spread over various geographic locations while maintaining the self-governance of each facility. The system's intelligent pattern recognition abilities allow for the dynamic fine-tuning of detection criteria, accommodating evolving threat scenarios. Thus, these adaptive frameworks significantly reduce the need for human intervention while preserving visibility, which is crucial for securing critical infrastructure.
\bibliographystyle{unsrt}  
\bibliography{ref}  

\begin{thebibliography}{10}

\bibitem{hong2014detection}
Junho Hong, Chen-Ching Liu, and Manimaran Govindarasu.
\newblock Detection of cyber intrusions using network-based multicast messages for substation automation.
\newblock In {\em ISGT 2014}, pages 1--5. IEEE, 2014.

\bibitem{sun2018cyber}
Chih-Che Sun, Adam Hahn, and Chen-Ching Liu.
\newblock Cyber security of a power grid: State-of-the-art.
\newblock {\em International Journal of Electrical Power \& Energy Systems}, 99:45--56, 2018.

\bibitem{hong2022automated}
Junho Hong, Tai-Jin Song, Hyojong Lee, and Aydin Zaboli.
\newblock Automated cybersecurity tester for \uppercase{IEC} 61850-based digital substations.
\newblock {\em Energies}, 15(21):7833, 2022.

\bibitem{10339874}
Silvio~E. Quincozes et~al.
\newblock Ereno: A framework for generating realistic \uppercase{IEC}–61850 intrusion detection datasets for smart grids.
\newblock {\em IEEE Transactions on Dependable and Secure Computing}, pages 1--15, 2023.

\bibitem{zaboli2025advanced}
Aydin Zaboli, Yong-Hwa Kim, and Junho Hong.
\newblock An advanced generative \uppercase{AI}-based anomaly detection in iec61850-based communication messages in smart grids.
\newblock {\em IEEE Access}, 2025.

\bibitem{beg2023review}
Omar~A Beg et~al.
\newblock A review of \uppercase{AI}-based cyber-attack detection and mitigation in microgrids.
\newblock {\em Energies}, 16(22):7644, 2023.

\bibitem{hong2017intelligent}
Junho Hong and Chen-Ching Liu.
\newblock Intelligent electronic devices with collaborative intrusion detection systems.
\newblock {\em IEEE Transactions on Smart Grid}, 10(1):271--281, 2017.

\bibitem{chen2016modeling}
Ying Chen, Junho Hong, and Chen-Ching Liu.
\newblock Modeling of intrusion and defense for assessment of cyber security at power substations.
\newblock {\em IEEE Transactions on Smart Grid}, 9(4):2541--2552, 2016.

\bibitem{OpenAIChatGPT}
OpenAI.
\newblock Chatgpt.
\newblock \url{https://openai.com/chatgpt}.
\newblock Accessed: Feb. 2024.

\bibitem{anthropic}
{Anthropic}.
\newblock Anthropic \uppercase{C}laude \uppercase{P}ro.
\newblock \url{https://www.anthropic.com}, 2023.
\newblock Accessed: 2024-11-10.

\bibitem{Copilot}
{Microsoft Corporation}.
\newblock Microsoft \uppercase{C}opilot \uppercase{AI}.
\newblock \url{https://copilot.microsoft.com/}, 2023.
\newblock Accessed: Feb. 1, 2024.

\bibitem{smolin2024gencoder}
Mikhail Smolin.
\newblock Gencoder: A generative \uppercase{AI}-based adaptive intra-vehicle intrusion detection system.
\newblock {\em IEEE Access}, 2024.

\bibitem{zaboli2024chatgpt}
Aydin Zaboli, Seong~Lok Choi, Tai-Jin Song, and Junho Hong.
\newblock Chatgpt and other large language models for cybersecurity of smart grid applications.
\newblock In {\em 2024 IEEE Power \& Energy Society General Meeting (PESGM)}, pages 1--5. IEEE, 2024.

\bibitem{zaboli2024leveraging}
Aydin Zaboli, Seong~Lok Choi, and Junho Hong.
\newblock Leveraging conversational generative \uppercase{AI} for anomaly detection in digital substations.
\newblock {\em arXiv preprint arXiv:2411.16692}, 2024.

\bibitem{gill2023chatgpt}
Sukhpal~Singh Gill and Rupinder Kaur.
\newblock Chat\uppercase{GPT}: Vision and challenges.
\newblock {\em Internet of Things and Cyber-Physical Systems}, 3:262--271, 2023.

\bibitem{ten2011anomaly}
Chee-Wooi Ten, Junho Hong, and Chen-Ching Liu.
\newblock Anomaly detection for cybersecurity of the substations.
\newblock {\em IEEE Transactions on Smart Grid}, 2(4):865--873, 2011.

\bibitem{choi2020multi}
In-Sun Choi, Junho Hong, and Tae-Wan Kim.
\newblock Multi-agent based cyber attack detection and mitigation for distribution automation system.
\newblock {\em IEEE Access}, 8:183495--183504, 2020.

\bibitem{kreimel2020anomaly}
Philipp Kreimel et~al.
\newblock Anomaly detection in substation networks.
\newblock {\em Journal of Information Security and Applications}, 54:102527, 2020.

\bibitem{wang2022anomaly}
Xuelei Wang et~al.
\newblock Anomaly detection for insider attacks from untrusted intelligent electronic devices in substation automation systems.
\newblock {\em IEEE Access}, 10:6629--6649, 2022.

\bibitem{alvee2021ransomware}
Syed~RB Alvee, Bohyun Ahn, Taesic Kim, Ying Su, Young-Woo Youn, and Myung-Hyo Ryu.
\newblock Ransomware attack modeling and artificial intelligence-based ransomware detection for digital substations.
\newblock In {\em 2021 6th IEEE Workshop on the Electronic Grid (eGRID)}, pages 01--05. IEEE, 2021.

\bibitem{panthi2020anomaly}
Manikant Panthi.
\newblock Anomaly detection in smart grids using machine learning techniques.
\newblock In {\em 2020 First International Conference on Power, Control and Computing Technologies (ICPC2T)}, pages 220--222. IEEE, 2020.

\bibitem{ankitdeshpandey2020development}
Ankitdeshpandey and R~Karthi.
\newblock Development of intrusion detection system using deep learning for classifying attacks in power systems.
\newblock In {\em Soft Computing: Theories and Applications: Proceedings of SoCTA 2019}, pages 755--766. Springer, 2020.

\bibitem{eynawi2024machine}
Ahmad Eynawi, Aneeqa Mumrez, Ghada Elbez, and Veit Hagenmeyer.
\newblock Machine learning-based feature selection for intrusion detection systems in \uppercase{IEC} 61850-based digital substations.
\newblock In {\em 2024 IEEE International Conference on Communications, Control, and Computing Technologies for Smart Grids (SmartGridComm)}, pages 1--7. IEEE, 2024.

\bibitem{quincozes2022feature}
Vagner~E Quincozes, Silvio~E Quincozes, C{\'e}lio Albuquerque, Diego Passos, and Daniel Moss{\'e}.
\newblock Feature extraction for intrusion detection in \uppercase{IEC}-61850 communication networks.
\newblock In {\em 2022 6th Cyber security in networking conference (CSNet)}, pages 1--7. IEEE, 2022.

\bibitem{jay2023deception}
Devika Jay.
\newblock Deception technology based intrusion protection and detection mechanism for digital substations: a game theoretical approach.
\newblock {\em IEEE Access}, 11:53301--53314, 2023.

\bibitem{bhattacharya2024ml}
Souradeep Bhattacharya, Nazmus Saqib, and Manimaran Govindarasu.
\newblock \uppercase{ML}-based anomaly detection system for \uppercase{IEC} 61850 communication in substations.
\newblock In {\em 2024 IEEE Power \& Energy Society General Meeting (PESGM)}, pages 1--5. IEEE, 2024.

\bibitem{upadhyay2020gradient}
Darshana Upadhyay, Jaume Manero, Marzia Zaman, and Srinivas Sampalli.
\newblock Gradient boosting feature selection with machine learning classifiers for intrusion detection on power grids.
\newblock {\em IEEE Transactions on Network and Service Management}, 18(1):1104--1116, 2020.

\bibitem{zhu2020intrusion}
Ruoxi Zhu, Chen-Ching Liu, Junho Hong, and Jiankang Wang.
\newblock Intrusion detection against mms-based measurement attacks at digital substations.
\newblock {\em IEEE Access}, 9:1240--1249, 2020.

\bibitem{ustun2021machine}
Taha~Selim Ustun et~al.
\newblock Machine learning-based intrusion detection for achieving cybersecurity in smart grids using \uppercase{IEC} 61850 \uppercase{GOOSE} messages.
\newblock {\em Symmetry}, 13(5):826, 2021.

\bibitem{yuan2023data}
Lixiang Yuan, Siyang Yu, Zhibang Yang, Mingxing Duan, and Kenli Li.
\newblock A data balancing approach based on generative adversarial network.
\newblock {\em Future Generation Computer Systems}, 141:768--776, 2023.

\bibitem{dromard2020study}
Juliette Dromard and Philippe Owezarski.
\newblock Study and evaluation of unsupervised algorithms used in network anomaly detection.
\newblock In {\em Proceedings of the Future Technologies Conference (FTC) 2019: Volume 2}, pages 397--416. Springer, 2020.

\bibitem{lin2019dynamic}
Peng Lin, Kejiang Ye, and Cheng-Zhong Xu.
\newblock Dynamic network anomaly detection system by using deep learning techniques.
\newblock In {\em Cloud Computing--CLOUD 2019: 12th International Conference, Held as Part of the Services Conference Federation, SCF 2019, San Diego, CA, USA, June 25--30, 2019, Proceedings 12}, pages 161--176. Springer, 2019.

\bibitem{mbona2022detecting}
Innocent Mbona and Jan~HP Eloff.
\newblock Detecting zero-day intrusion attacks using semi-supervised machine learning approaches.
\newblock {\em IEEE Access}, 10:69822--69838, 2022.

\bibitem{yaacoub2020cyber}
Jean-Paul~A Yaacoub, Ola Salman, Hassan~N Noura, Nesrine Kaaniche, Ali Chehab, and Mohamad Malli.
\newblock Cyber-physical systems security: Limitations, issues and future trends.
\newblock {\em Microprocessors and microsystems}, 77:103201, 2020.

\bibitem{fu2022deep}
Yanfang Fu, Yishuai Du, Zijian Cao, Qiang Li, and Wei Xiang.
\newblock A deep learning model for network intrusion detection with imbalanced data.
\newblock {\em Electronics}, 11(6):898, 2022.

\bibitem{boukerche2020outlier}
Azzedine Boukerche, Lining Zheng, and Omar Alfandi.
\newblock Outlier detection: Methods, models, and classification.
\newblock {\em ACM Computing Surveys (CSUR)}, 53(3):1--37, 2020.

\bibitem{dairi2023semi}
Abdelkader Dairi, Fouzi Harrou, Benamar Bouyeddou, Sidi-Mohammed Senouci, and Ying Sun.
\newblock Semi-supervised deep learning-driven anomaly detection schemes for cyber-attack detection in smart grids.
\newblock In {\em Power systems cybersecurity: Methods, concepts, and best practices}, pages 265--295. Springer, 2023.

\bibitem{lopez2022substation}
Jose~Antonio Lopez, I{\~n}aki Angulo, and Saturnino Martinez.
\newblock Substation-aware. an intrusion detection system for the iec 61850 protocol.
\newblock In {\em Proceedings of the 17th International Conference on Availability, Reliability and Security}, pages 1--7, 2022.

\bibitem{sahani2023machine}
Nitasha Sahani, Ruoxi Zhu, Jin-Hee Cho, and Chen-Ching Liu.
\newblock Machine learning-based intrusion detection for smart grid computing: A survey.
\newblock {\em ACM Transactions on Cyber-Physical Systems}, 7(2):1--31, 2023.

\bibitem{anwar2021comparison}
Mahwish Anwar, Anton Borg, and Lars Lundberg.
\newblock A comparison of unsupervised learning algorithms for intrusion detection in iec 104 scada protocol.
\newblock In {\em 2021 International Conference on Machine Learning and Cybernetics (ICMLC)}, pages 1--8. IEEE, 2021.

\bibitem{singh2021cyber}
Vivek~Kumar Singh and Manimaran Govindarasu.
\newblock A cyber-physical anomaly detection for wide-area protection using machine learning.
\newblock {\em IEEE Transactions on Smart Grid}, 12(4):3514--3526, 2021.

\bibitem{khaw2020deep}
Yew~Meng Khaw, Amir~Abiri Jahromi, Mohammadreza~FM Arani, Scott Sanner, Deepa Kundur, and Marthe Kassouf.
\newblock A deep learning-based cyberattack detection system for transmission protective relays.
\newblock {\em IEEE Transactions on Smart Grid}, 12(3):2554--2565, 2020.

\bibitem{lim2024future}
Willone Lim, Kelvin Sheng~Chek Yong, Bee~Theng Lau, and Colin Choon~Lin Tan.
\newblock Future of generative adversarial networks (gan) for anomaly detection in network security: A review.
\newblock {\em Computers \& Security}, 139:103733, 2024.

\bibitem{sauber2022use}
Rick Sauber-Cole and Taghi~M Khoshgoftaar.
\newblock The use of generative adversarial networks to alleviate class imbalance in tabular data: a survey.
\newblock {\em Journal of Big Data}, 9(1):98, 2022.

\bibitem{manzoor2025detecting}
Faizan Manzoor, Vanshaj Khattar, Akila Herath, Clifton Black, Matthew~C Nielsen, Junho Hong, Chen-Ching Liu, and Ming Jin.
\newblock Detecting zero-day attacks in digital substations via in-context learning.
\newblock {\em arXiv preprint arXiv:2501.16453}, 2025.

\bibitem{manzoor2024zero}
Faizan Manzoor, Vanshaj Khattar, Chen-Ching Liu, and Ming Jin.
\newblock Zero-day attack detection in digital substations using in-context learning.
\newblock In {\em 2024 IEEE International Conference on Communications, Control, and Computing Technologies for Smart Grids (SmartGridComm)}, pages 220--225. IEEE, 2024.

\bibitem{2024_3C_causalprompt}
Tung-Wei Lin, Vanshaj Khattar, Yuxuan Huang, Junho Hong, Ruoxi Jia, Chen-Ching Liu, Alberto Sangiovanni-Vincentelli, and Ming Jin.
\newblock \uppercase{CausalPrompt}: Enhancing \uppercase{LLM}s with weakly supervised causal reasoning for robust performance in non-language tasks.
\newblock In {\em ICLR Workshop: Tackling Climate Change with Machine Learning}, 2024.

\bibitem{quincozes2023ereno}
Silvio~Ereno Quincozes, C{\'e}lio Albuquerque, Diego Passos, and Daniel Moss{\'e}.
\newblock Ereno: A framework for generating realistic iec--61850 intrusion detection datasets for smart grids.
\newblock {\em IEEE Transactions on Dependable and Secure Computing}, 21(4):3851--3865, 2023.

\bibitem{hong2021implementation}
Junho Hong et~al.
\newblock Implementation of secure sampled value (\uppercase{S}e\uppercase{SV}) messages in substation automation system.
\newblock {\em IEEE Transactions on Power Delivery}, 37(1):405--414, 2021.

\bibitem{biswas2019synthesized}
Partha~P Biswas, Heng~Chuan Tan, Qingbo Zhu, Yuan Li, Daisuke Mashima, and Binbin Chen.
\newblock A synthesized dataset for cybersecurity study of iec 61850 based substation.
\newblock In {\em 2019 IEEE International Conference on Communications, Control, and Computing Technologies for Smart Grids (SmartGridComm)}, pages 1--7. IEEE, 2019.

\bibitem{hong2014integrated}
Junho Hong, Chen-Ching Liu, and Manimaran Govindarasu.
\newblock Integrated anomaly detection for cyber security of the substations.
\newblock {\em IEEE Transactions on Smart Grid}, 5(4):1643--1653, 2014.

\bibitem{bhuyan2013network}
Monowar~H Bhuyan, Dhruba~Kumar Bhattacharyya, and Jugal~K Kalita.
\newblock Network anomaly detection: methods, systems and tools.
\newblock {\em Ieee communications surveys \& tutorials}, 16(1):303--336, 2013.

\bibitem{garcia2014empirical}
Sebastian Garcia, Martin Grill, Jan Stiborek, and Alejandro Zunino.
\newblock An empirical comparison of botnet detection methods.
\newblock {\em computers \& security}, 45:100--123, 2014.

\bibitem{buda2018systematic}
Mateusz Buda, Atsuto Maki, and Maciej~A Mazurowski.
\newblock A systematic study of the class imbalance problem in convolutional neural networks.
\newblock {\em Neural networks}, 106:249--259, 2018.

\bibitem{yu2022pricing}
Mingkai Yu, Jianxiao Wang, Jie Yan, Lin Chen, Yang Yu, Gengyin Li, and Ming Zhou.
\newblock Pricing information in smart grids: A quality-based data valuation paradigm.
\newblock {\em IEEE Transactions on Smart Grid}, 13(5):3735--3747, 2022.

\bibitem{yang2015cybersecurity}
Yi~Yang, HT~Jiang, Kieran McLaughlin, L~Gao, YB~Yuan, W~Huang, and Sakir Sezer.
\newblock Cybersecurity test-bed for iec 61850 based smart substations.
\newblock In {\em 2015 IEEE Power \& Energy Society General Meeting}, pages 1--5. IEEE, 2015.

\bibitem{zeng2024divtod}
Weihao Zeng, Dayuan Fu, Keqing He, Yejie Wang, Yukai Xu, and Weiran Xu.
\newblock \uppercase{D}iv\uppercase{TOD}: Unleashing the power of \uppercase{LLM}s for diversifying task-oriented dialogue representations.
\newblock {\em arXiv preprint arXiv:2404.00557}, 2024.

\bibitem{paek2024enhancing}
Ellie~S Paek, Talyn Fan, James~D Finch, and Jinho~D Choi.
\newblock Enhancing task-oriented dialogue systems through synchronous multi-party interaction and multi-group virtual simulation.
\newblock {\em Information}, 15(9):580, 2024.

\bibitem{su2024large}
Jing Su, Chufeng Jiang, Xin Jin, Yuxin Qiao, Tingsong Xiao, Hongda Ma, Rong Wei, Zhi Jing, Jiajun Xu, and Junhong Lin.
\newblock Large language models for forecasting and anomaly detection: A systematic literature review.
\newblock {\em arXiv preprint arXiv:2402.10350}, 2024.

\bibitem{huang2024design}
Yu~Huang and Liangyuan Su.
\newblock Design of intrusion detection and response mechanism for power grid scada based on improved lstm and fnn.
\newblock {\em IEEE Access}, 2024.

\bibitem{usmani2022review}
Usman~Ahmad Usmani, Ari Happonen, and Junzo Watada.
\newblock A review of unsupervised machine learning frameworks for anomaly detection in industrial applications.
\newblock In {\em Science and Information Conference}, pages 158--189. Springer, 2022.

\bibitem{meng2020time}
Chao Meng, Xue~Song Jiang, Xiu~Mei Wei, and Tao Wei.
\newblock A time convolutional network based outlier detection for multidimensional time series in cyber-physical-social systems.
\newblock {\em IEEE Access}, 8:74933--74942, 2020.

\bibitem{jindal2016decision}
Anish Jindal, Amit Dua, Kuljeet Kaur, Mukesh Singh, Neeraj Kumar, and Sukumar Mishra.
\newblock Decision tree and svm-based data analytics for theft detection in smart grid.
\newblock {\em IEEE Transactions on Industrial Informatics}, 12(3):1005--1016, 2016.

\bibitem{vaswani2017attention}
Ashish Vaswani, Noam Shazeer, Niki Parmar, Jakob Uszkoreit, Llion Jones, Aidan~N Gomez, {\L}ukasz Kaiser, and Illia Polosukhin.
\newblock Attention is all you need.
\newblock {\em Advances in neural information processing systems}, 30, 2017.

\bibitem{de2022general}
Isaac~Mart{\'\i}n De~Diego et~al.
\newblock General performance score for classification problems.
\newblock {\em Applied Intelligence}, 52(10):12049--12063, 2022.

\end{thebibliography}

\end{document}